\pdfoutput=1
\documentclass{pasj02}
\usepackage[switch,mathlines]{lineno}
\usepackage{url}
\usepackage{amsmath}
\usepackage{bm}

\newcommand{\rev}[1]{\textcolor{black}{#1}}

\Received{$\langle$reception date$\rangle$}
\Accepted{$\langle$acception date$\rangle$}
\Published{$\langle$publication date$\rangle$}

\begin{document}

\title{
%Constraints of gas-to-halo mass relation by 
%cross correlations of 2MASS galaxies with cosmic dispersion measures along 133 localized fast radio bursts
Cross-correlating galaxies and cosmic dispersion measures:
Constraints on the gas-to-halo mass relation from 2MASS galaxies and 133 localized fast radio bursts
}

\author{Masato \textsc{Shirasaki}\altaffilmark{1,2}\altemailmark, Ryuichi \textsc{Takahashi}\altaffilmark{3}, Ken \textsc{Osato}\altaffilmark{4-6}, and Kunihito \textsc{Ioka}\altaffilmark{7}}% \orcid{0000-0000-0000-0000}}%

\altaffiltext{1}{National Astronomical Observatory of Japan (NAOJ), National Institutes of Natural Sciences, Tokyo 181-8588, Japan}
\altaffiltext{2}{The Institute of Statistical Mathematics, Tokyo 190-8562, Japan}
%\altaffiltext{3}{RIKEN Center for Advanced Intelligence Project, Tokyo 103-0027, Japan}
\altaffiltext{3}{Faculty of Science and Technology, Hirosaki University, Aomori 036-8560, Japan}
\altaffiltext{4}{Center for Frontier Science, Chiba University, Chiba 263-8522, Japan}
\altaffiltext{5}{Department of Physics, Graduate School of Science, Chiba University, Chiba 263-8522, Japan}
\altaffiltext{6}{Kavli Institute for the Physics and Mathematics of the Universe (WPI), The University of Tokyo
Institutes for Advanced Study, The University of Tokyo, Chiba 277-8583, Japan}
\altaffiltext{7}{Yukawa Institute for Theoretical Physics, Kyoto University, Kyoto 606-8502, Japan}
\email{masato.shirasaki@nao.ac.jp}

\KeyWords{methods: statistical --- intergalactic medium --- large-scale structure of universe}

\maketitle

\begin{abstract}
We conduct a cross-correlation analysis between large-scale structures traced by the Two Micron All Sky Survey (2MASS) galaxy catalog and the cosmic dispersion measures of 133 localized fast radio bursts (FRBs). The cross-correlation signal is measured as a function of the comoving separation $R$ between 2MASS galaxies and background FRB sightlines, making full use of the available redshift information for both datasets. Our measurements are consistent with a null detection over the range $0.01 < R\, [h^{-1}\mathrm{Mpc}] < 1$.
Using a halo-based model in which free-electron density profiles are drawn from the hydrodynamical simulation IllustrisTNG-300 (TNG300), we show that the null signal at $R \sim 0.01\, h^{-1}\mathrm{Mpc}$ is inconsistent with the TNG300 prediction. This discrepancy indicates that the hot-gas mass fraction in halos with masses of 
$10^{12-13}\, M_\odot$ hosting 2MASS galaxies must be lower than that predicted by TNG300. A simple phenomenological modification of the TNG300 model suggests that the hot-gas mass fraction in halos of $10^{12-13}\, M_\odot$ should be below $\sim 10\%$ of the global baryon fraction in the nearby universe, implying the need for stronger feedback in this mass range. Our constraints are consistent with those inferred from X-ray emission and Sunyaev-Zel'dovich measurements in galaxies, while providing a direct estimate of the hot-gas mass fraction that does not rely on electron-temperature measurements. These results demonstrate that galaxy-FRB cross correlations offer a powerful probe of feedback processes in galaxy formation.
\end{abstract}

%\pagewiselinenumbers

\section{Introduction}

Fast radio bursts (FRBs) are radio transients with
their typical pulse duration being several milliseconds (see \cite{2019A&ARv..27....4P} for a review).
Although their physical origin is yet unclear \citep{2023RvMP...95c5005Z}, 
FRBs have attracted much attention as a next-generation probe of cosmology. The arrival time of FRB pulses is known to be dispersed due to light propagation through intervening electron plasma. 
The frequency dependence in the arrival time $t_\mathrm{arr}$ is written as \begin{equation}
t_\mathrm{arr}(\nu)=\frac{e^2}{2\pi m_e c} \frac{\mathrm{DM}}{\nu^2} = 4.14\, \mathrm{ms} \left(\frac{\mathrm{DM}}{\mathrm{pc}/\mathrm{cc}}\right)\left(\frac{\nu}{1\mathrm{GHz}}\right)^{-2},
\end{equation}
where $e$ is the electron charge, $m_e$ is the electron mass, $c$ is the speed of light,
$\nu$ is the observed frequency, 
and $\mathrm{DM}$ is called the dispersion measure, equaling the column density of free electrons along the line-of-sight to the source.
Hence, measurements of delays in FRB-pulse arrivals provide a unique means of mapping free electrons in the universe, which is an important pillar of baryonic components in the concordance cosmological model.

Baryons are known to occupy only a $\sim5\%$ of the energy-density budget in the present-day universe (e.g.~\cite{2020A&A...641A...6P}), but their feedback plays an essential role in shaping our observable universe.
For instance, stellar winds and supernovae explosions can regulate star-forming processes across galactic scales (e.g.~\cite{1986ApJ...303...39D}), while active galactic nuclei are an important contributor to
foundation of scaling relations in galaxy clusters (e.g.~\cite{2021Univ....7..142E})
as well as termination of star formations at massive galaxies (e.g.~\cite{2012ARA&A..50..455F}).
Furthermore, modern cosmology faces a major challenge in quantifying the baryonic feedback and its impact on statistical analyses of cosmic large-scale structures as precise datasets become available \citep{2019OJAp....2E...4C}.

% S8 tension
Current understanding of baryonic feedback in large-scale structures remains limited. A series of Stage-III galaxy imaging surveys has demonstrated precise estimation of cosmological parameters using weak lensing effects %sourced
caused by low-redshift ($z \sim 0.5\text{--}2$) matter distributions alone (e.g.~\cite{2021A&A...645A.104A, 2022PhRvD.105b3514A, 2022PhRvD.105b3515S, 2023PhRvD.108l3519D, 2023PhRvD.108l3518L}). However, the accuracy of these parameter estimates depends critically on the details of baryonic feedback modeling (e.g.~\cite{2022MNRAS.516.5355A, 2023MNRAS.518.5340C, 2023A&A...678A.109A, 2024MNRAS.534..655B, 2025PhRvD.111f3509T, 2025A&A...703L...3B}).

% hydrodynamical simulations and subgrid physics
Hydrodynamical simulations covering cosmological volumes are among the most essential approaches for predicting baryonic feedback effects in large-scale structures. Despite their success, these simulations face a fundamental challenge in spatial resolution: it is impractical to capture the enormous dynamic range of relevant physical scales, from sub-$\mathrm{pc}$ to $O(1)\,\mathrm{Gpc}$, within a single simulation. A common practice is to introduce subgrid recipes to reduce computational costs in the highest-density environments (e.g.~\cite{2025arXiv250206954V} for a review). However, this phenomenological methodology must be validated against multiple observations.

% observational tests so far
Most cosmological hydrodynamical simulations calibrate their subgrid recipes using observations of stellar components across a wide range of redshifts, as well as intracluster media at low redshifts. 
Observational information about gaseous matter in and around galaxy-sized objects is highly sought to further improve the realism of these simulations. 
In this context, statistical analyses using large samples of FRBs have been proposed as a way to overcome current limitations, offering valuable insights into baryonic feedback effects. These include one-point probability functions of dispersion measures \citep{2024ApJ...967...32M, 2025ApJ...993..162Z, 2025ApJ...989...81S, 2025MNRAS.540..289G, 2025ApJ...983...46M, 2025arXiv250717742R}, autocorrelation functions of FRB dispersion measures \citep{2021MNRAS.502.2615T, 2022JCAP...04..046N}, cross-correlation functions with tracers of large-scale structures \citep{2022MNRAS.512.1730S, 2023arXiv230909766R, 2025ApJ...993L..27H, 2025arXiv250608932W, 2025arXiv250919514L, 2025arXiv251102155T}, and various combinations of these approaches \citep{2017PhRvD..95h3012S, 2025arXiv250905866S}.

In this paper, we perform a cross-correlation analysis between a galaxy catalog with spectroscopic redshift information and 133 localized FRBs.
Our FRB sample is introduced in the companion paper \citep{2025arXiv251102155T}, which compiles publicly available FRBs with precise localizations.
We propose a novel estimator for real-space cross correlations, incorporating a robust background selection based on redshifts.
This estimator is designed to 
%be free from contamination 
\rev{significantly mitigate contamination}
by Galactic and FRB-host dispersion measures, 
%providing a clean probe of 
\rev{enabling a relatively clean probe of}
the free-electron density  at the redshifts of foreground galaxies.
Our measurements can be readily interpreted using a standard halo-based modeling framework \citep{2002PhR...372....1C}, widely adopted in studies of large-scale structure. Using well-defined cross-correlation measurements and an efficient modeling approach, we aim to place independent constraints on gas-to-halo mass relations in the local universe at a redshift of $z \sim 0.03$.
\rev{Given the current observational uncertainties, our measurements are primarily sensitive to the amplitude of the electron-density profile, and thus mainly constrain the total gas fraction within halos rather than the detailed radial distribution.}

It is worth noting that \citet{2025ApJ...993L..27H} also conducted a similar cross-correlation analysis, but there are notable differences between the two studies.
We use spectroscopic galaxy redshifts for accurate measurements, whereas \citet{2025ApJ...993L..27H} relied on multiple galaxy catalogs with photometric redshifts.
Their analysis is based on 61 localized FRBs, but their foreground galaxy sample is larger than ours.
We pay particular attention to measuring cross correlations at scales down to 
$\sim 0.01 \,\mathrm{Mpc}$, as small-scale information is crucial for constraining baryonic physics in group-sized objects -- an aspect not addressed in \citet{2025ApJ...993L..27H}.
We expect that the photometric-redshift uncertainties of their foreground galaxy sample limit their ability to probe such small scales in \citet{2025ApJ...993L..27H}.

The remainder of this paper is organized as follows:
Section~\ref{sec:data} describes the dataset used for the cross-correlation measurements.
Section~\ref{sec:measurement} summarizes our estimator and the methodology for assessing statistical and systematic uncertainties.
Section~\ref{sec:model} introduces the halo-model framework used to generate predictions for comparison with our measurements.
Our main results are presented in Section~\ref{sec:results}, and we conclude in Section~\ref{sec:conc}.

Throughout this paper, we assume a spatially flat $\Lambda$CDM cosmology with
$\Omega_\mathrm{m} = 1-\Omega_\Lambda = 0.315$,
$\Omega_\mathrm{b} = 0.049$,
$h = 0.674$,
$n_s = 0.965$, and
$\sigma_8 = 0.811$,
consistent with recent measurements of cosmic microwave background anisotropies by the Planck satellite \citep{2020A&A...641A...6P}.
We adopt the spherical-overdensity mass convention, $M_\Delta = 4\pi r_\Delta^3 \rho_{\mathrm{crit},z} \Delta / 3$, where $r_\Delta$ is the halo radius within which the mean density equals $\Delta$ times the critical density $\rho_{\mathrm{crit},z}$ at redshift $z$. Unless otherwise stated, we use $M \equiv M_{200}$ as our fiducial choice.
Conversions between different spherical-overdensity masses follow \citet{2003ApJ...584..702H}.

%\clearpage

\section{Data}\label{sec:data}

\subsection{2MASS galaxy catalog}

We use the galaxy catalog from the 2MASS Redshift Survey (2MRS; \cite{2012ApJS..199...26H}), a spectroscopic follow-up of the Two Micron All Sky Survey \citep{2006AJ....131.1163S}.
\citet{2018MNRAS.473.4318A} showed that the redshift distribution of 2MASS galaxies is well described by the following functional form:
\begin{equation}
p_\mathrm{2MASS}(z) = \frac{n}{z_0 \Gamma[(m+1)/n]}\left(\frac{z}{z_0}\right)^m \exp\left[-\left(\frac{z}{z_0}\right)^m\right], \label{eq:2MASS_pz}
\end{equation}
where $\Gamma(x)$ is the Gamma function, and the parameters $m = 1.31$, $n = 1.64$, and $z_0 = 0.0266$ are inferred in \citet{2018MNRAS.473.4318A}.
The 2MRS provides a nearly volume-limited sample up to $z \simeq 0.02$ with high redshift completeness for galaxies with $K_s \le 11.75$.
Clustering analyses in \citet{2018MNRAS.473.4318A} also constrain the halo occupation distributions (HOD) of 2MASS galaxies with good precision.
We rely on their HOD model to infer typical host halo masses of 2MASS galaxies when interpreting our cross-correlation measurements. Details of the HOD are given in Section~\ref{sec:model}.

In this paper, we account for redshift completeness\footnote{This is 
the fraction of objects in the 2MASS photometric sample for which reliable redshift measurements have been obtained.}
using information from \citet{2011MNRAS.416.2840L} and the HEALPix\footnote{\url{https://healpix.sourceforge.io/}} package \citep{2005ApJ...622..759G}.
First, we compute the average completeness in a HEALPix map with resolution $\mathrm{NSIDE}=32$, referring to completeness at $K_s \le 11.5$ from the catalog in \citet{2011MNRAS.416.2840L}.
We find that HEALPix pixels with average completeness greater than 0.5 cover about 
$\sim 80\%$ of the sky.
Each 2MASS galaxy is then assigned a weight equal to the average completeness of its corresponding HEALPix pixel.
To avoid regions near the Galactic plane in our cross-correlation analysis, we set the weight to zero for galaxies at $|b| < 5^\circ$ for $30^\circ < l < 330^\circ$ and $|b| < 10^\circ$ otherwise, 
where $b$ and $l$ denote Galactic latitude and longitude, respectively.
We also generate 1000 random catalogs using the completeness map and uniform sampling of random points on a sphere.
Each random catalog contains 43,533 entries, matching the actual 2MASS catalog.
These random catalogs are used to subtract the mean offset of dispersion measures from our cross-correlation measurements.
Additionally, we divide the galaxy catalog into 40 subregions using the dsigma\footnote{\url{https://dsigma.readthedocs.io/en/latest/index.html}} package \citep{2022MNRAS.515.4722H}.
These subregions with similar area allow us to estimate the covariance matrix of our measurements using the standard delete-one jackknife method commonly employed in galaxy analyses \citep{2009MNRAS.396...19N, 2017MNRAS.470.3476S, 2017MNRAS.471.3827S}.

\rev{We have also tested the robustness of the jackknife covariance against the choice of the number of subregions. By varying the number of subregions from 30 to 50 around our fiducial choice of 40, we find that the inferred signal-to-noise ratios fluctuate at a level consistent with the expected statistical uncertainty given the number of degrees of freedom. This indicates that our main conclusions are not strongly sensitive to the choice of jackknife configuration, although the precise $p$-values should be interpreted with some caution.}

\begin{figure}
 \begin{center}
  \includegraphics[width=80mm]{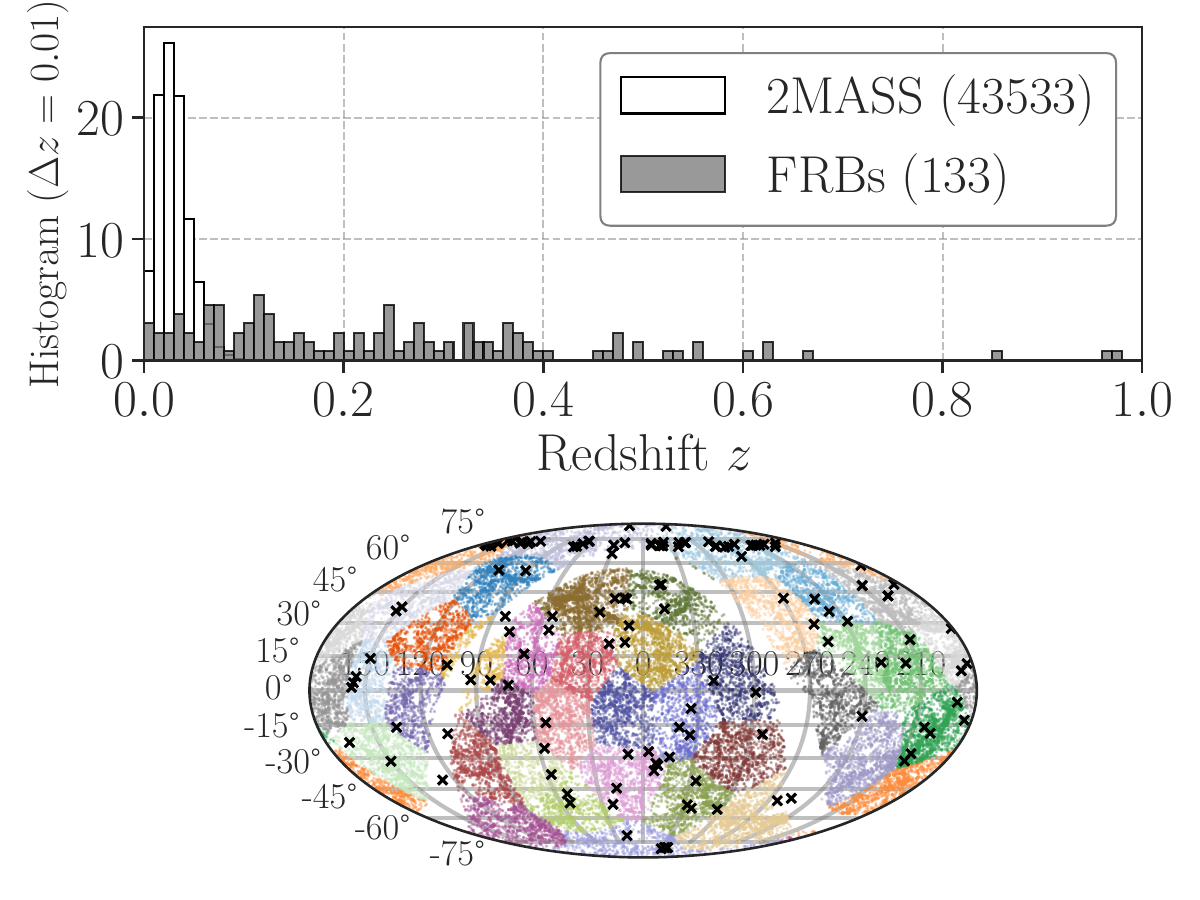}
 \end{center}
 \caption{Redshift and angular distributions of our dataset. 
 The upper panel summarizes the histogram of 2MASS-galaxy and FRB redshifts with the bin width of $\Delta z=0.01$,
 while colored points and black cross symbols in the lower panel represent the angular positions of 2MASS galaxies and 133 localized FRBs, respectively. We adopt the equatorial coordinate system in the lower panel. Note that the different colors in the lower panel highlight 40 jackknife subdivisions of the 2MASS galaxy catalog. Those subregions are used for the covariance estimation in our cross-correlation measurements. \label{fig:catalog}
 {Alt text: Graphs and data on spatial information of galaxies and fast radio bursts.}}
\end{figure}

The upper panel of Fig.~\ref{fig:catalog} shows the redshift distribution of 2MASS galaxies, while the lower panel illustrates the large-scale structure traced by these galaxies across the full sky.
Different colors in the lower panel indicate the subregions used for jackknife-based covariance estimation.

\subsection{Localized fast radio bursts}

We use the sample of localized FRBs summarized in \citet{2025arXiv251102155T}.
The authors compiled 133 FRBs whose host galaxies were identified through follow-up observations reported in the literature 
(see Table~6 in \cite{2025arXiv251102155T} for the full list).
They then inferred the cosmic dispersion measures\footnote{An observed dispersion measure can be expressed as the sum of three components: contributions from intervening large-scale structures (LSS) between the FRB and the observer, the FRB host galaxy, and the Milky Way (MW). We define the “cosmic” dispersion measure as the sum of the LSS and host-galaxy components.} for individual FRBs using a likelihood analysis based on a mixture of three log-normal distributions of dispersion measures.
Their likelihood analysis adopted two different models for the MW contribution \citep{2002astro.ph..7156C, 2017ApJ...835...29Y, 2020ApJ...888..105Y} and two different redshift evolutions for the host-galaxy component.
For the redshift dependence of host-galaxy dispersion measures, the authors used a common model assuming a power-law evolution with $(1+z)^{-1}$ \citep{2003ApJ...598L..79I, 2014PhRvD..89j7303Z}, while also exploring a freely varying power-law index motivated by recent theoretical studies \citep{2020AcA....70...87J, 2020ApJ...900..170Z, 2023MNRAS.518..539M, 2024A&A...690A..47K, 2024arXiv240308611T, 2025OJAp....8E.127R}.
In the end, their inferred cosmic dispersion measures for the localized FRBs are categorized into four classes: N1, Nbeta, Y1, and Ybeta.
We use these four estimates in our cross-correlation measurements to assess systematic uncertainties arising from the component separation of observed dispersion measures.

It is important to note that the FRB angular positions and redshifts in \citet{2025arXiv251102155T} are precisely determined through follow-up observations.
This accurate localization enables us to measure the cross-correlation between cosmic dispersion measures and the large-scale structure traced by 2MASS galaxies as a function of comoving separation.
Furthermore, the secure redshift estimates allow us to select background FRBs relative to individual 2MASS galaxies, eliminating potential contamination from host-galaxy dispersion measures in our cross-correlation analysis.
This setup is crucial for estimating the average free-electron density profile around 2MASS galaxies alone.
It is worth noting that the previous work \citep{2025ApJ...993L..27H} did not pay attention to this background selection.
We describe the details of our cross-correlation measurement procedure in Section~\ref{sec:measurement}.
The redshift distribution of the FRB sample is shown in the upper panel of Fig.~\ref{fig:catalog}, while crosses in the lower panel indicate the FRB locations.

\section{Cross-correlation measurements}\label{sec:measurement}

The cross-correlation function between galaxies and FRB dispersion measures is defined as
\begin{equation}
\xi_\mathrm{gd}(R) = \langle \delta_\mathrm{g}({\bm \theta}_g, z_g) 
\mathrm{DM}({\bm \theta}_s, z_s)\rangle, \label{eq:def_2pcf}
\end{equation}
where $\delta_\mathrm{g}({\bm \theta}_g, z_g)$ is the overdensity field of the galaxy 
with its angular position ${\bm \theta}_g$ and redshift $z_g$,
and $\mathrm{DM}({\bm \theta}_s, z_s)$ represents the dispersion measure coming from 
the FRB with its angular position ${\bm \theta}_s$ and redshift $z_s$.
In Eq.~(\ref{eq:def_2pcf}), we define a comoving separation $R$ by using known redshifts of galaxies as 
%$R = \chi(z_g)|{\bm \theta}_g - {\bm \theta}_s|$,
$R = \chi(z_g) \Delta \theta_{gs}$,
where $\chi(z)$ is the comoving distance to redshift $z$
and $\Delta \theta_{gs}$ is the angular separation between galaxy and FRB in each pair.
This definition of $\xi_\mathrm{gd}$ is intended to measure the average free-electron density profile around the foreground galaxy sample, borrowing the concept from galaxy–galaxy lensing measurements (e.g.~\cite{2015IAUS..311...86M, 2020A&ARv..28....7U} for reviews).
When background FRBs can be securely selected relative to the galaxies based on their redshifts, the cross-correlation in Eq.~(\ref{eq:def_2pcf}) 
%arises solely from 
\rev{predominantly reflects}
the number-density perturbations of free electrons at the galaxy redshifts $z_g$.
This enables a robust measurement of the average hot-gas mass associated with the foreground galaxies 
%without contamination from 
\rev{while reducing the impact of} FRB-host dispersion measures, providing valuable insights into baryonic feedback processes in the universe.
\rev{We note, however, that this observable probes a weighted projection of the electron density field rather than the gas mass itself, and the inference of gas mass therefore depends on assumptions about the gas distribution. The sensitivity to different spatial scales and the impact of profile assumptions are discussed in detail in Sections~\ref{sec:model} and \ref{sec:results}.}

\subsection{Estimator}

In practice, we can estimate the cross-correlation of $\xi_\mathrm{gd}$ from a discrete set of positions and redshifts of galaxies and FRBs as
\begin{equation}
\hat{\xi}_\mathrm{gd}(R_i) = \frac{\sum_{gs \in R_i} w_g w_s \Delta \mathrm{DM}_s}{\sum_{gs \in R_i} w_g w_s} - (\mathrm{signal}\, \mathrm{around}\, \mathrm{random})|_{R_i}, \label{eq:est_2pcf}
\end{equation}
where 
the summation ``$gs$'' run over all galaxy-FRB pairs that lie in the $i$-th radial bin $R_i \equiv \chi(z_g) \Delta \theta_{gs}$;
$w_g$ represents the weight of the $g$-th galaxy accounting for the redshift completeness in the 2MRS;
$w_s$ is the weight of the $s$-th FRB introduced to downweight the contribution from FRBs with large uncertainties in estimating MW dispersion measures;
we define $\Delta \mathrm{DM}_s$ as residual dispersion measure of the $s$-th FRB;
\begin{equation}
\Delta \mathrm{DM}_s \equiv \mathrm{DM}_{\mathrm{obs},s} - \mathrm{DM}_{\mathrm{MW},s} - \int_0^{z_s}\frac{c\mathrm{d}z}{H(z)} \bar{n}_\mathrm{e} (1+z), \label{eq:Delta_DM_s}
\end{equation}
where $\mathrm{DM}_{\mathrm{obs},s}$ and $\mathrm{DM}_{\mathrm{MW},s}$ are the observed and MW dispersion measures for the $s$-th FRB, $H(z)$ is the Hubble parameter at $z$, and $\bar{n}_\mathrm{e}$ represents the mean comoving number density of free electrons.
We use an esimate of $\bar{n}_\mathrm{e}$ inferred from the likelihood analysis in \citet{2025arXiv251102155T},
while it depends on the detail of component separations.
Note that we impose a selection of background FRBs in the computation of Eq.~(\ref{eq:est_2pcf}) with $z_s \ge z_g + \Delta z$, where $\Delta z$ is a parameter to minimize interlopers in our analysis.
In Eq.~(\ref{eq:Delta_DM_s}), we do not subtract the FRB-host component to make our analysis insensitive to the choice of FRB models.
We set $\Delta z= 0.05$ throughout this paper.
Note that a more conservative choice 
with $\Delta z=0.1$ does not affect our measurements (but increases the statistical errors).
The weight $w_s$ is set to an inverse variance by using the variance of MW dispersion measure 
on an FRB-by-FRB basis.
Finally, the second term on the right-hand side of Eq.~(\ref{eq:est_2pcf})
denotes the signal around a set of random positions, 
which is measured by replacing foreground galaxies with random points. 
We need to subtract this random signal to evaluate an excess from a uniform contribution of dispersion measures to all radial bins. 

% binning
In the measurements, we employ a logarithmic binning with 10 bins ranging from 
$R=0.01\, h^{-1}\mathrm{Mpc}$ to $R=100\, h^{-1}\mathrm{Mpc}$ for our analysis.
% statistical error (jackknife, random shuffling)
The covariance matrix of our cross-correlation measurements is estimated with 
the 40 jackknife samples for our fiducial choice. 
For comparison, we also estimate the covariance matrix of $\hat{\xi}_\mathrm{gd}$ 
by producing 300 realizations of $\Delta \mathrm{DM}$ with random shuffling of dispersion measures (we randomly shuffle $\Delta\mathrm{DM}_s$ but fix each FRB coordinate).
Note that the jackknife errors include sample-variance contributions but those cannot be applied to the data at large $R$ beyond the size of jackknife subregions.
The random-shuffling errors can be used as a robust estimate of shot noises 
over a wide range of $R$.
Hence, we decide to use the data for scientific analyses only when its jackknife error is greater than the random-shuffling counterpart. 

\begin{figure}
 \begin{center}
  \includegraphics[width=80mm]{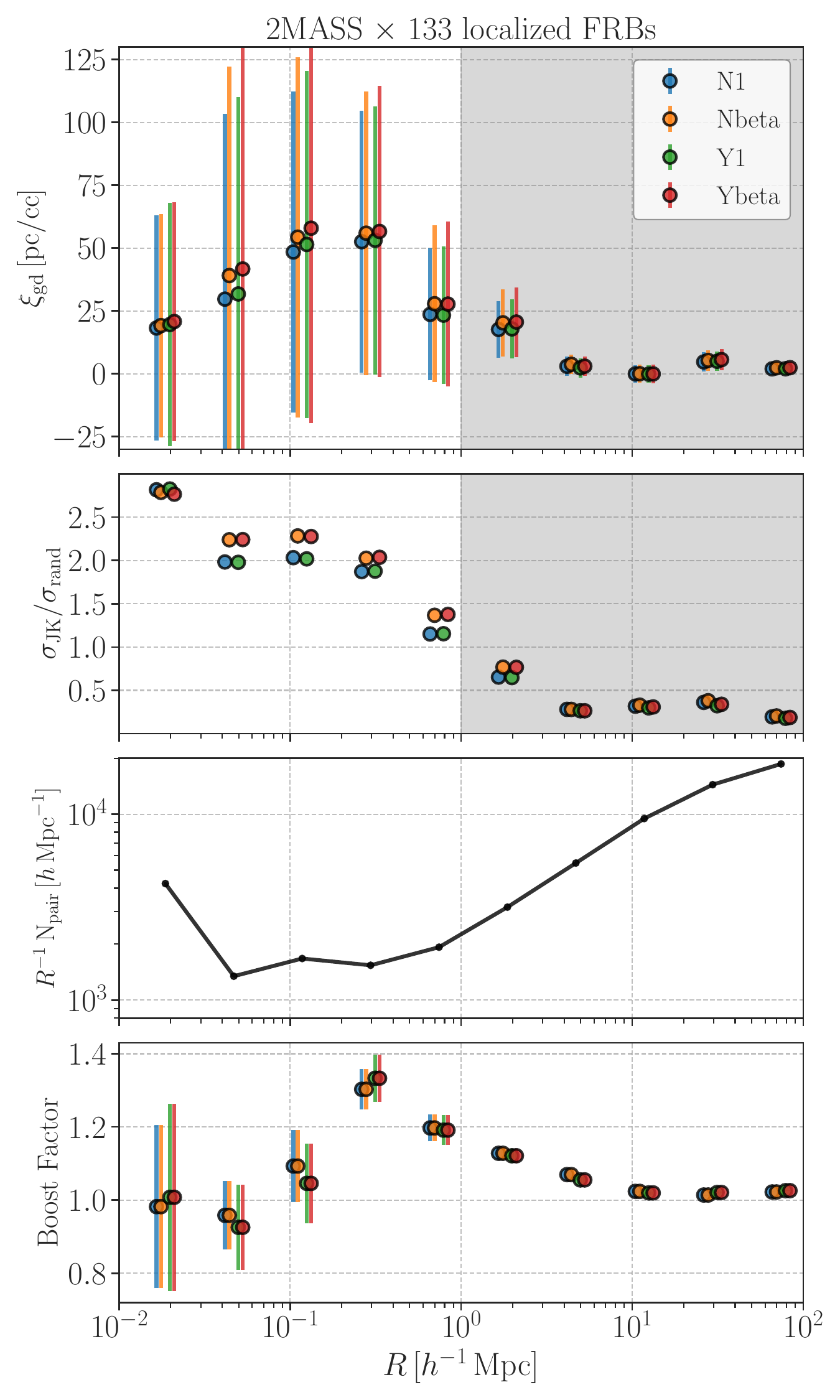}
 \end{center}
 \caption{A summary of our cross-correlation measurements.
 The top, middle upper, middle lower, and bottom panels show the cross-correlation functions between 2MASS and FRB dispersion measures, the ratio of the jackknife errors to the random-shuffling shot noises,
 the number of galaxy-FRB pairs at each radial bin,
 and the boost factor (a diagonistic quantity for secure selections of background FRBs), respectively.
 Different colored points in this figure highlight that our measurements rely on the four different estimates of dispersion measures $\Delta \mathrm{DM}_s$.
 The gray filled regions in the top two panels represent the jackknife-based estimate of statistical errors at $R>1\, h^{-1}\mathrm{Mpc}$ is not appropriate for scientific analyses.
 Note that our analysis will be subject to misidentification of background FRBs 
 or/and observational selection effects of FRBs
 when the boost factor significantly deviates from unity.
 We describe the boost factor and its role in Section~\ref{subsec:boost}.
 \label{fig:summary_cross_correlations}
 {Alt text: Graphs and data on our cross-correlation measurements.}}
\end{figure}

Figure~\ref{fig:summary_cross_correlations} summarizes our measurements.
The top panel shows the cross-correlation functions with the jackknife errors.
We have four different measurements there as four different estimates of 
$\Delta \mathrm{DM}$ are available.
The middle upper panel represents the ratio of the jackknife error to 
the random-shuffling counterpart.
The gray filled region in the upper two panels highlights the data with 
its jackknife errors of $\xi_\mathrm{gd}$ smaller than the random-shuffling ones.
To make our analysis conservative, we avoid to use the data in the gray region for comparison with theoretical models, leaving 5 bins in the range of $0.01<R\, [h^{-1}\mathrm{Mpc}]<1$ in the end.
For reference, we also show the number of pairs in each radial bin at the middle lower panel.
To be specific, 339 and 2224 galaxy-FRB pairs are available at $R < 0.1\, h^{-1}\mathrm{Mpc}$ and $R < 1\, h^{-1}\mathrm{Mpc}$, respectively.

\subsection{Boost factor}\label{subsec:boost}

Although our analysis introduces the non-zero parameter $\Delta z$ to ensure secure background selection, it remains possible that some FRBs are physically associated with their foreground galaxies.
Misidentification of host galaxies can occur, particularly for non-repeating FRBs.
Additionally, current observations may miss certain bursts if they experience significant pulse broadening due to multi-path propagation in inhomogeneous plasma (e.g.~\cite{2021arXiv211207639S, 2022ApJ...927...35C, 2023AJ....165..152M, 2025arXiv250906721S}).
At face value, the top panel of Fig.~\ref{fig:summary_cross_correlations} shows a trend of decreasing $\xi_\mathrm{gd}$ as the galaxy-FRB separation becomes smaller.
This trend is consistent with the expectation that FRB pulse scattering times increase in denser environments (e.g., galactic centers; see also \cite{2016arXiv160505890C, 2019MNRAS.483..971V, 2022ApJ...934...71O}), suggesting that our cross-correlation measurements may be affected by scattering effects.

To test whether our background selection is impacted by these possibilities, we use the ratio of the number of galaxy-FRB pairs to the number of random-FRB pairs.
This ratio, known as the boost factor in galaxy-galaxy lensing analyses \citep{2004MNRAS.353..529H, 2005MNRAS.361.1287M}, allows us to assess the validity of our background selection in the cross-correlation measurements.
Accounting for weights, we define the boost factor as
\begin{equation}
B(R_i) = \left(\sum_{gs} w_g w_s / \sum_{gs} w_g\right) / \left(\sum_{rs} w_r w_s / \sum_{rs} w_r\right),
\end{equation}
where 
the summation ``$gs$'' and ``$rs$'' run over all galaxy-FRB pairs and random-FRB pairs that lie in the $i$-th radial bin $R_i$, respectively.
Note that $B > 1$ is expected if some FRBs are physically associated with foreground galaxies, while $B < 1$ suggests that some FRBs may be missed due to scattering effects on their pulses.

The bottom panel of Fig.~\ref{fig:summary_cross_correlations} summarizes our boost factor measurements.
We estimate the errors using 300 bootstrap realizations of random catalogs.
We find that boost factors at $R \lesssim 0.1\,h^{-1}\mathrm{Mpc}$ are consistent with unity, indicating that our FRB sample is unaffected by scattering effects.
At $R \gtrsim 0.3\,h^{-1}\mathrm{Mpc}$, we observe a moderate deviation from unity, but this has minimal impact on our cross-correlation measurements given the large statistical uncertainties.
%The radius of $R=0.3\, h^{-1}\mathrm{Mpc}$ roughly corresponds to $0.1$ degrees for the median redshift of the 2MASS galaxies, probably relating to locarization errors in current FRB observations.
\rev{The radius of $R=0.3, h^{-1}\mathrm{Mpc}$ roughly corresponds to $\sim 0.1$ degrees at the median redshift of the 2MASS galaxies. While this scale is somewhat larger than the naive statistical localization uncertainties of current FRB samples 
(e.g.~\cite{2025ApJS..280....6C}), it may reflect additional systematic uncertainties arising from host-galaxy misidentification or associations with faint galaxies at different redshifts. In any case, such uncertainties do not have a significant impact on our main results.}
We therefore conclude that our background selection is sufficiently robust for the cross-correlation analysis at this stage.

\subsection{Statistical significance}

The detection significance of the cross correlation function is characterized as the signal-to-noise ratio, which is defined by
\begin{equation}
\left(\mathrm{S/N}\right)^2 = \sum_{ij} \hat{\xi}_\mathrm{gd}(R_i) \, \mathrm{Cov}^{-1}_{ij} \, \hat{\xi}_\mathrm{gd}(R_j), 
\end{equation}
where $\hat{\xi}_\mathrm{gd}(R_i)$ represents the cross correlation function at the $i$-th radial bin as measured in Eq.~(\ref{eq:est_2pcf}) and $\mathrm{Cov}$ is the jackknife covariance matrix for our measurements.
In the range of $0.01<R\, [h^{-1}\mathrm{Mpc}] < 1$, we find that 
$\left(\mathrm{S/N}\right)^2 = 9.44, 8.44, 11.8$, and $10.4$ for the estimates with N1, Nbeta, Y1, and Ybeta, respectively\footnote{The corresponding $p$-values are computed as
0.092, 0.13, 0.037, and 0.064 for 5 degrees of freedom.}.
Since we have 5 degrees of freedom in our measurements, those values of $\left(\mathrm{S/N}\right)^2$ are consistent with a null detection at a 97\% confidence level. 
%\KO{[KO: Corresponding $p$-values (or $x\text{-}\sigma$) may be useful.]}
For the sake of completeness, we provide numerical values of $\hat{\xi}_\mathrm{gd}(R_i)$ and its covariance matrix in Appendix~1. 

% chi^2 against a null hypothesis in the range of 0.01<R<1 with 5 bins
% N1 9.440924e+00
% Nbeta 8.438763e+00
% Y1 1.183342e+01
% Ybeta 1.043224e+01

\section{Model}\label{sec:model}

We here describe a theoretical model to be compared with our measurements.
The angular number density of foreground galaxies can be decomposed into
\begin{equation}
n_\mathrm{g}({\bm \theta}_g) = \bar{n}_\mathrm{g}\left[1+\delta_\mathrm{g}({\bm \theta}_g)\right],
\end{equation}
where $\bar{n}_\mathrm{g}$ is the average angular number density and it is also computed as
\begin{equation}
\bar{n}_\mathrm{g} = \int \mathrm{d}z \, p(z)\, \frac{\mathrm{d}\chi}{\mathrm{d}z}\, \chi^2 \, \bar{n}_\mathrm{g,3D}(z),
\end{equation}
where $p(z)$ is the selection function of galaxies in redshifts 
(given by Eq.~(\ref{eq:2MASS_pz}) in our case),
and we introduce $\bar{n}_\mathrm{g,3D}(z)$ to be the three-dimensional average number density of the galaxies at redshift $z$ in the comoving frame.
%Note that the selection function is normalized to as $\int \mathrm{d}\chi\, f_\mathrm{g}(\chi)=1$.
%In our case with the 2MASS redshift distribution given in Eq.~(\ref{eq:2MASS_pz}), it holds that $f_\mathrm{g}(\chi) = p_\mathrm{2MASS}(z[\chi])(\mathrm{d}\chi/\mathrm{d}z)^{-1}$.
The overdensity field $\delta_\mathrm{g}$ is then written in terms of its three-dimensional counterpart as
\begin{equation}
\delta_\mathrm{g}({\bm \theta}_g) = \frac{1}{\bar{n}_\mathrm{g}}\int \mathrm{d}z \, p(z)\, \frac{\mathrm{d}\chi}{\mathrm{d}z} \chi^2 \, \bar{n}_\mathrm{g,3D}(z)\, \delta_\mathrm{g,3D}({\bf x}).
\end{equation}
Here we denote the three-dimensional overdensity field as 
$\delta_\mathrm{g,3D}({\bf x})$ at the position of ${\bf x}=(\chi, \chi {\bm \theta})$.

Dispersion measures from intervening plasma between sources at redshift $z_s$ and an observer are formally computed as
\begin{equation}
\mathrm{DM}({\bm \theta}_s, z_s) = \int_0^{z_s}\mathrm{d}\chi \, (1+z[\chi])\, n_\mathrm{e}(\chi, \chi{\bm \theta}_s),
\end{equation}
where $n_\mathrm{e}$ represents the comoving number density of free electrons.
Suppose we can surely select the background sources with redshift information, the cross correlation function can be computed as (e.g.~\cite{2022MNRAS.512.1730S})
\begin{align}
\xi_\mathrm{gd}(R) &= \frac{1}{\bar{n}_\mathrm{g}}\int \mathrm{d}z\, p(z)\, \frac{\mathrm{d}\chi}{\mathrm{d}z}\chi^2 \, (1+z) \, \bar{n}_\mathrm{g,3D}(z) \nonumber \\
& \quad \quad \quad \quad \quad \quad  \quad \times \int \frac{k\mathrm{d}k}{2\pi}P_\mathrm{gd}(k;\chi)J_0(kR), \label{eq:model_2pcf} \\
P_\mathrm{gd}(k;\chi) &\equiv \int \mathrm{d}^3 r\, \langle \delta_\mathrm{g,3D}({\bf x}) n_\mathrm{e}({\bf x}+{\bf r})\rangle|_{\chi} \exp(-{\bf k}\cdot{\bf r}),
\end{align}
where 
$\langle \cdots\rangle|_{\chi}$ is evaluated as an emsenble average at the comoving radial distance $\chi$,
$J_0(x)$ is the zeroth-order Bessel function,
and we have adopted the Limber approximation.
It is worth noting that the observed dispersion measures should contain the contributions from the Galaxy as well as near-soure plasma, but our estimator of Eq.~(\ref{eq:est_2pcf}) is designed to be 
%immune to 
\rev{minimally affected by} those contaminations.
Hence, we can focus on the correlation between free electrons and galaxies sharing the same large-scale structures.

We use a halo model \citep{2002PhR...372....1C} to compute the cross power spectrum of $P_\mathrm{gd}$. In the halo model, we can decompose the power spectrum into two terms as
\begin{equation}
P_\mathrm{gd}(k) = P_\mathrm{gd,1h}(k) + P_\mathrm{gd,2h}(k), \label{eq:pk_halomodel}
\end{equation}
where the first term in the right hand side is called the one-halo term 
arising from two-point correlations inside single halos,
while the second one is the two-halo term corresponding to the correlations between different halos.
In the following, we specify our model ingredients to compute $P_\mathrm{gd}$ as well as the three-dimensional average number density of galaxies $\bar{n}_\mathrm{g,3D}(z)$.

\subsection{Halo occupation distribution of 2MASS galaxies}

On the modeling of the 2MASS galaxies, we adopt an HOD to relate the galaxy number density with its host-halo counterpart;
\begin{equation}
n_\mathrm{g,3D}({\bf x}) = \sum_{i} \left[\langle N_\mathrm{cen}|M_i \rangle \delta^{(3)}_D({\bf x}-{\bf x}_i) +  \langle N_\mathrm{sat}|M_i \rangle U_\mathrm{sat}({\bf x}-{\bf x}_i)\right], \label{eq:def_hod}
\end{equation}
where the summation run over all dark-matter halos, 
${\bf x}_i$ represents the position of the $i$-th halo, 
$\delta^{(3)}_\mathrm{D}({\bf r})$ is the three-dimensional delta function,
and $U_\mathrm{sat}({\bf r})$ describes the spacial distribution of satellite galaxies in single host halos. 
In Eq.~(\ref{eq:def_hod}), the terms of $\langle N_\mathrm{cen}|M\rangle$
and $\langle N_\mathrm{sat}|M\rangle$ are called the halo occupation number, 
meaning the number of central and satellite galaxies in single host halos with their masses of $M$, respectively.

For the halo occupation number and the satellite profile of $U_\mathrm{sat}$,
we use the model as constrained by the power-spectrum analysis in \citet{2018MNRAS.473.4318A}. Those are given by
\begin{align}
\langle N_\mathrm{cen}|M\rangle &= \frac{1}{2}\left[1+\mathrm{erf}\left(\frac{\log M - \log M_\mathrm{min}}{\sigma_{\log M}}\right)\right], \label{eq:Ncen} \\
\langle N_\mathrm{sat}|M\rangle &= \left(\frac{M-M_\mathrm{min}}{M_1}\right)^{\alpha}\Theta(M-M_\mathrm{min}), \label{eq:Nsat} \\
U_\mathrm{sat}(r) &\propto \frac{\Theta(r_\mathrm{max,g}-r)}{(r/r_\mathrm{s,g})(1+r/r_{\mathrm{s,g}})^2}, \label{eq:Usat}
\end{align}
where 
$\Theta(x)$ is the Heaviside step function.
For the halo occupation number, we set the parameters of 
$\log(M_\mathrm{min}/M_\odot) = 11.84$, 
$\sigma_{\log M}=0.15$, $\log(M_\mathrm{1}/M_\odot) = 11.98$, 
and $\alpha=0.849$, while
$r_\mathrm{s,g} = 0.62 \, c^{-1}_{200} r_\mathrm{200}$,
and $r_\mathrm{max,g}=6.9 \, r_\mathrm{200}$ are adopted in Eq.~(\ref{eq:Usat})
with $c_\mathrm{200}$ being the halo concentration parameter.
Throughout this paper, we use the model of $c_\mathrm{200}$ in \citet{2008MNRAS.390L..64D}.
Note that we normalize the satellite profile such that $\int 4\pi r^2 \mathrm{d} r \, U_\mathrm{sat}(r) = 1$.
\rev{We have verified that adopting the HOD parameter sets corresponding to Run 1–3 in Table 2 of \citet{2018MNRAS.473.4318A} changes our model predictions by $\lesssim 5\%$, 
indicating that our results are not strongly sensitive to the choice of HOD parameters.}

In the end, the average number density of galaxies is then computed as 
\begin{equation}
\bar{n}_\mathrm{g,3D}(z) = \int \mathrm{d}M\, n_\mathrm{h}(M,z)\, 
\left[\langle N_\mathrm{cen}|M\rangle + \langle N_\mathrm{sat}|M\rangle\right],
\end{equation}
where $n_\mathrm{h}(M,z)$ is the halo mass function (i.e.~the number density of halos with $M$ at redshift $z$). We use the fitting function of 
$n_\mathrm{h}(M,z)$ as developed in \citet{2008ApJ...688..709T}.

\subsection{Free electrons around dark matter haloes}

In the halo-model framework, we require the electron number density profile around individual halos to compute Eq.~(\ref{eq:pk_halomodel}).
For our fiducial choice, we extract spherically symmetric profiles of electron number density around dark matter halos from the IllustrisTNG300 simulation \citep{2018MNRAS.475..676S,2018MNRAS.475..648P,2018MNRAS.475..624N,2018MNRAS.480.5113M,2018MNRAS.477.1206N}.
We use the particle snapshot and associated halo catalog at $z = 0$ from the TNG300-1 run.
This run employs a box size of $205\,h^{-1}\mathrm{Mpc}$ per side and includes $2500^3$ gas cells and $2500^3$ dark matter particles, enabling us to study free-electron number density across a wide range of halo masses.

For our analysis, we adopt logarithmic halo-mass bins, dividing the range $11 < \log M_{500}/(h^{-1}M_\odot) < 14.5$ into 35 bins.
Within each mass bin, we compute the average free-electron number density profile as a function of $r/r_\mathrm{500}$, where $r$ is the halo-centric radius.
Profiles are evaluated in 30 logarithmic bins spanning $0.01 < r/r_\mathrm{500} < 10$.
Even in the most massive bin, we find 14 halos, ensuring that our stacking analysis is robust against statistical fluctuations.
To obtain the free-electron abundance from gas cells, we follow the approach in \citet{2025arXiv250707090K}.
In IllustrisTNG, the electron number density is computed based on ionization equilibrium of primordial gas \citep{1996ApJS..105...19K}. The computed number density is valid for non-star-forming gas cells. However, for star-forming cells, each gas cell is composed of cold and hot interstellar media \citep{2003MNRAS.339..289S}, and only the latter contributes to the free electron number density. Here, we simply ignore the star-forming gas cells because the cold interstellar medium accounts for roughly $90\%$ of the total density.
%We ignore the free-electron density of star-forming gas and compute the electron density of non-star-forming gas under the assumption of full ionization.
This approach provides a lower limit on the electron density profile in the TNG300-1 simulation.
However, the prediction already conflicts with our measurements at small scales (see Fig.~\ref{fig:model_vs_data}); including star-forming gas cells would only worsen the discrepancy, making the model even less compatible with observations.

Given the tables of the electron number density profile $n_\mathrm{e,h}(r, M)$,
we then compute the halo-model power spectrum of Eq.~(\ref{eq:pk_halomodel}) as
(e.g.~see \cite{2013MNRAS.430..725V} for a similar expression in the context of galaxy-galaxy lensing)
\begin{align}
P_\mathrm{gd,1h}(k) &= \int \mathrm{d}M\, n_\mathrm{h}(M)\, \tilde{U}_\mathrm{g}(k)\, \tilde{n}_\mathrm{e,h}(k,M), \label{eq:pk_1h} \\
P_\mathrm{gd,2h}(k) &= P_\mathrm{L}(k) \left\{\int \mathrm{d}M\, b_\mathrm{h}(M)\, n_\mathrm{h}(M)\, \tilde{n}_\mathrm{e,h}(k,M) \right\} \nonumber \\
& \quad \quad \quad \times \left\{\int \mathrm{d}M\, b_\mathrm{h}(M)\, n_\mathrm{h}(M)\, \tilde{U}_\mathrm{g}(k)\right\}, \label{eq:pk_2h} \\
\tilde{U}_\mathrm{g}(k) &= \langle N_\mathrm{cen}|M\rangle + \langle N_\mathrm{sat}|M\rangle \, \tilde{U}_\mathrm{sat}(k),
\end{align}
where the quantity with tilde represents the Fourier transform of its correspondence, 
$b_\mathrm{h}(M)$ is the linear bias of halos with $M$,
and $P_\mathrm{L}(k)$ is the linear power spectrum of cosmic mass density.
Throughout this paper, we adopt the halo bias model $b_\mathrm{h}(M)$ from \citet{2010ApJ...724..878T} and use the Boltzmann solver CAMB \citep{2000ApJ...538..473L} to compute $P_\mathrm{L}(k)$.
In Eqs.~(\ref{eq:pk_1h}) and (\ref{eq:pk_2h}), the lower and upper limits of the integral are set to $10^{10.5} \, h^{-1}M_\odot$ and $10^{16}\,h^{-1}M_\odot$, respectively.
For halo masses outside the range $11 < \log M_{500}/(h^{-1}M_\odot) < 14.5$, we assume that the profile shape can be approximated by that of the lowest or highest mass bin, while scaling the amplitude using a logarithmic extrapolation of the gas-to-halo mass relation.
Specifically, we compute the (hot) gas mass for a halo of mass $M$ as
\begin{equation}
M_\mathrm{gas}(M)=\int_0^{r_{500}} \mathrm{d}r\, 4\pi r^2 \, n_\mathrm{e,h}(r,M) \, m_\mathrm{p}\, \mu_\mathrm{e},
\end{equation}
where $m_\mathrm{p}$ is the proton mass,
$\mu_\mathrm{e}=(X_\mathrm{p}+Y_\mathrm{p}/2)^{-1}$ is the mean molecular weight of electrons, $X_\mathrm{p}$ and $Y_\mathrm{p}$ represent the primordial mass fractions of
hydrogen and helium, respectively. In this paper, we set $X_\mathrm{p} = 1-Y_\mathrm{p} = 0.76$.

\begin{figure}
 \begin{center}
  \includegraphics[width=80mm]{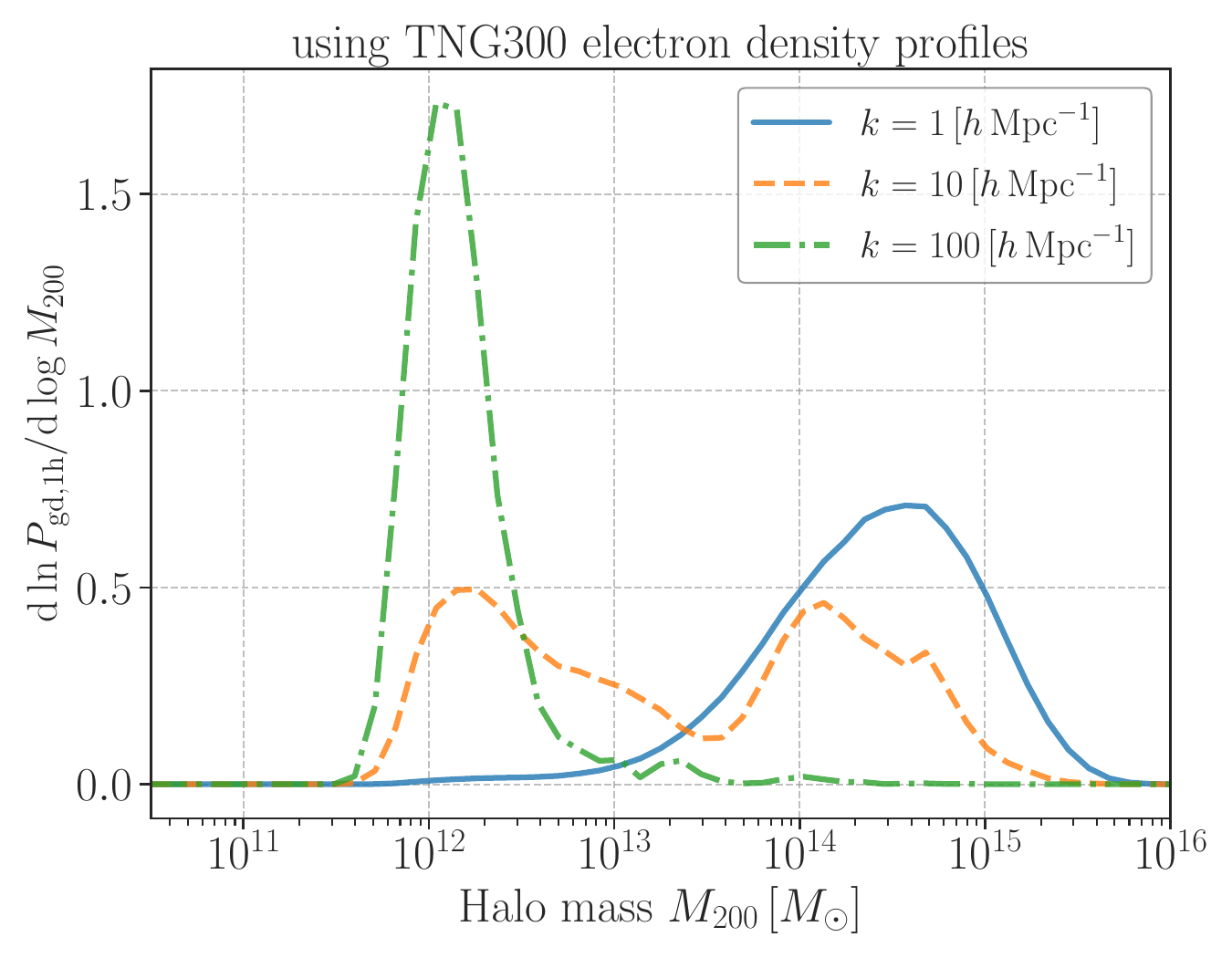}
 \end{center}
 \caption{The derivative of the one-halo cross power spectrum (defined in Eq.~\ref{eq:pk_1h}) with respect to halo masses. 
 In this figure, the blue solid, orange dashed, and green dashed-dotted lines represent the results for three scales of $1$, $10$, and $100\, h\mathrm{Mpc}^{-1}$, respectively. We here set the redshift to be 0.03 as an effective redshift of the 2MASS galaxies. \label{fig:halo_model_info}
 {Alt text: Three line graphs.}}
\end{figure}

Note that the contributions outside from $11 < \log M_{500}/(h^{-1}M_\odot)<14.5$ minimally affect our halo-model computation at $R\lesssim 0.1 h^{-1}\mathrm{Mpc}$, which is the main focus in this paper.
Figure~\ref{fig:halo_model_info} shows the derivatives of the one-halo term (Eq.~\ref{eq:pk_1h}) to the halo mass $M$ at three different scales of $k=1, 10$ and $100\, h\mathrm{Mpc}^{-1}$. 
At each fourier mode $k$, we compute the derivative as
\begin{equation}
\partial P_\mathrm{gd,1h}(k)/\partial M = n_\mathrm{h}(M) \, \tilde{U}_\mathrm{g}(k)\, \tilde{n}_\mathrm{e,h}(k,M),
\end{equation}
which is useful to see an effective halo mass in our halo-model predictions.
The figure highlights that the cross power spectrum of $P_\mathrm{gd}$ contains the information of cluster-scale halos at $k \sim 1\, \mathrm{Mpc}^{-1}$,
while smaller-scale information of $P_\mathrm{gd}$ can give valuable insights into less massive objects with $M\sim10^{12-13}\, M_\odot$.
\rev{Since Fourier modes at wavenumber $k$ predominantly contribute to the correlation function at real-space scales $R \sim 1/k$ (see Eq.~\ref{eq:model_2pcf}), these features directly map onto the signal at the corresponding radii in the cross-correlation function.}
The two-halo terms are also found to be subdominant at $R<1\, h^{-1}\mathrm{Mpc}$
in our analysis setup.

% UP TO HERE [2025.12.22]

%\clearpage

\section{Results}\label{sec:results}

\subsection{Comparison with model predictions}

\begin{figure}
 \begin{center}
  \includegraphics[width=80mm]{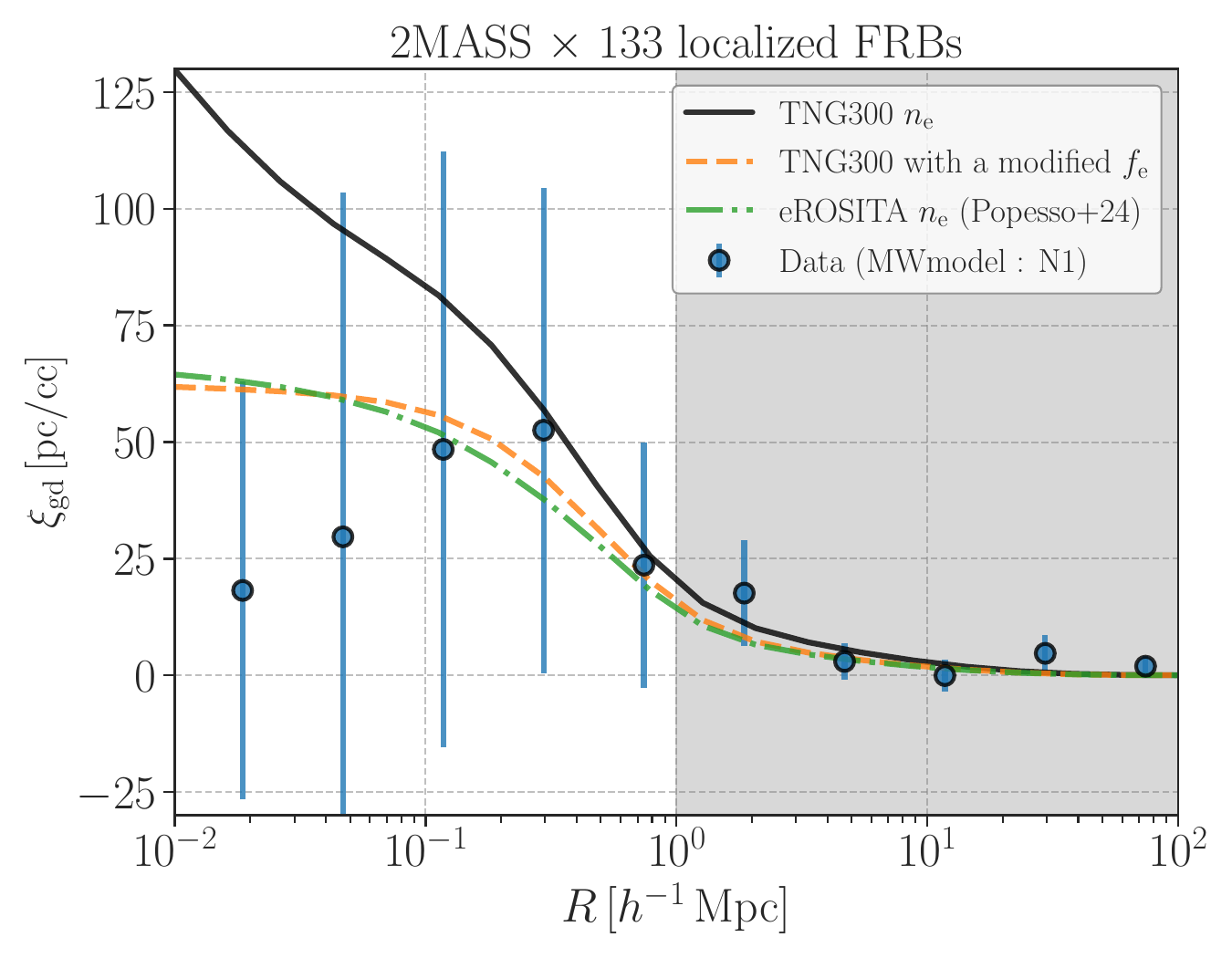}
 \end{center}
 \caption{Comparions of our measurement with several model predicitons.
 The blue points with error bars show the measured cross correlation functions beween the 2MASS galaxies and 133 localized FRBs, while the black solid line stands for our fiducial model prediciton based on the electron density profiles extracted from the TNG300-1 simulation. The orange dashed line represents the prediction based on the TNG300 outputs but we reduced the electron abundance $f_\mathrm{e}$ in halos with $M$ by a phenomenological function of $\tanh(M_{500}/10^{13.9} M_\odot)$. The green dashed-dotted line is an alternative to the TNG300 electron density profiles with observationally inferred ones in \citet{2024arXiv241116555P}. Note that the gray region in this figure highlights our covariance estimate of the cross correlation may be underestimated. \label{fig:model_vs_data}
 {Alt text: Graphs and data on the comparison with our cross-correlation measurements and theoretical predictions.}}
\end{figure}

We first compare our measurements of $\xi_\mathrm{gd}$ with 
model predictions based on the TNG-300 simulation.
For this purpose, we use the data vector based on FRB dispesion measures 
with the use of N1 template.
%Also, we apply a correction to mitigate possible interlopers in our background selection method by dividing the data vector to the boost factor at each radial bin.
Figure~\ref{fig:model_vs_data} summarizes the comparison, showcasing that
the TNG300-based prediction (the black line in the figure) is imcompatible with our measurement at $R \sim 0.01\, \mathrm{Mpc}$.
According to the information content discussed in Fig.~\ref{fig:halo_model_info}, 
we speculate this difference can be caused by an overpredicted hot gas mass 
at group-sized halos in the TNG-300 simulations.
To quantify the difference, we simply modify our halo-model predictions by reducing contributions from lower-mass halo as
\begin{equation}
n_\mathrm{e,h}(r, M) \rightarrow \tanh\left(M_{500}/M_\mathrm{cut}\right) n_\mathrm{e,h}(r,M),
\label{eq:tng_modified}
\end{equation}
where $M_\mathrm{cut}$ is a parameter in this phenomenological model. 
The orange dashed line in the figure shows the predictions with $M_\mathrm{cut} = 10^{13.9} M_\odot$, being in agreement with our measurements.

Another check is also performed with a different functional form in electron density profiles derived in \citet{2024arXiv241116555P}.
The model of \citet{2024arXiv241116555P} has been constructed from X-ray stacking analyses for optically-selected galaxy groups in the GAMA survey \citep{2011MNRAS.416.2640R} and X-ray data from eROSITA Science Verification data over the eROSITA Final Equatorial Depth Survey (eFEDS; \cite{2022A&A...661A...1B}).
They found the data can be reasonably fitted with an electron density profile as
\begin{align}
n_\mathrm{e, P24}(r) &= n_0 (r/r_c)^{-\alpha/2}\left[1+\left(r/r_c\right)^2\right]^{-3\beta/2+\alpha/4} \nonumber \\ 
& \quad \quad \quad \quad \quad \quad \times \left[1+\left(r/r_s\right)^3\right]^{-\epsilon/6}, \label{eq:ne_P24}
\end{align}
where the parameters of $n_0, \alpha, \beta, \epsilon, r_c$, and $r_s$ are all dependent of halo masses $M$. For those parameters, we interpolate the results in Table~1 in \citet{2024arXiv241116555P}.
It is worth noting that Eq.~(\ref{eq:ne_P24}) predicts the hot gas mass fraction $M_\mathrm{gas}/M \sim 0.1 \, \Omega_\mathrm{b}/\Omega_\mathrm{m}$ at $M\sim 10^{13} M_\odot$, while the fraction at the same mass range gets higher by a factor of $2-3$ in the TNG model.
The green dashed-dotted line in Fig.~\ref{fig:model_vs_data} shows that the model prediction based on Eq.~(\ref{eq:ne_P24}) is also consistent with our measurements, necessitating that the hot gas mass at $M\sim 10^{13} M_\odot$ should be of an order of $10\%$ of the global baryon fraction. 
\rev{Figure~\ref{fig:model_vs_data} also indicates that the amplitude of the gas density profile is the dominant factor in interpreting our measurements. This suggests that our analysis primarily constrains the total gas fraction, rather than being strongly sensitive to the detailed radial distribution of free electrons.}

\subsection{Implication to gas-to-halo mass relations}

\begin{figure*}
 \begin{center}
  \includegraphics[width=160mm]{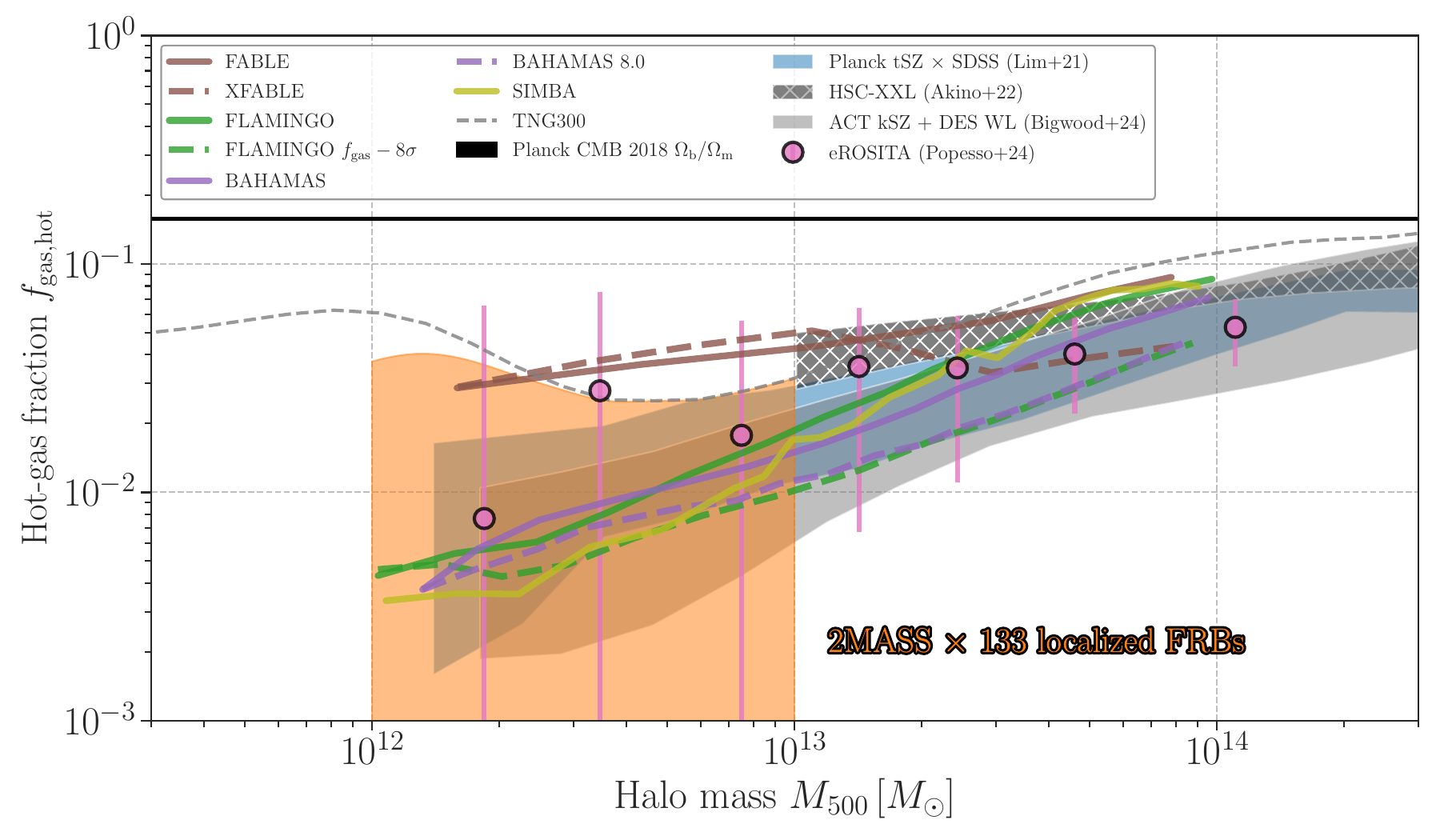}
 \end{center}
 \caption{The limit of gas-to-halo mass relations by our cross correlation measurements along with others. The orange filled region shows our constraint with a 95\% confidence level, while the cyan filled, black hatched, gray filled regions and pink points with error bars represent the observational limits based on \citet{2021MNRAS.504.5131L},
 \citet{2022PASJ...74..175A},
 \citet{2024MNRAS.534..655B},
 and \citet{2024arXiv241116555P}, respectively.
 For reference, we also plot the model predictions based on different hydrodynamical simulations; gray dashed line stands for the TNG300, 
 while brown, green, purple, yellow lines correspond to the predictions 
 for the FABLE and XFABLE suite \citep{2018MNRAS.479.5385H, 2025MNRAS.542.3206H},
 the FLAMINGO suite \citep{2023MNRAS.526.4978S, 2023MNRAS.526.6103K},
 the BAHAMAS suite \citep{2017MNRAS.465.2936M},
 and the SIMBA \citep{2019MNRAS.486.2827D}, respectively.
 The black solid horizontal line represents the global baryon fraction with the best-fit values inferred from the Planck observation \citep{2020A&A...641A...6P}.
 \label{fig:fgas_limit}
 {Alt text: Graphs and data on constraints of the gas-to-halo mass relation.}}
\end{figure*}

Using the null detection and the phenomenological model with Eq.~(\ref{eq:tng_modified}),
we can put a lower limit of $M_\mathrm{cut}$ as 
$M_\mathrm{cut} > 1\times 10^{12} M_\odot$ at a $95\%$ confidence level.
This is derived by a one-parameter likelihood analysis with the data in the range 
of $0.01<R\, [h^{-1}\mathrm{Mpc}]<0.2$, where the boost factors are consistent with unity.
In practice, we set the limit by solving the equations below;
\begin{align}
\Delta \chi^2 (M_\mathrm{cut}) &= \chi^2(M_\mathrm{cut}) -\min[\chi^2(M_\mathrm{cut})] = 4, \\
\chi^2(M_\mathrm{cut}) &= \sum_{ij}\Delta \xi_\mathrm{gd}(R_i|M_\mathrm{cut}) \, \mathrm{Cov}_{ij}^{-1} \,\Delta \xi_\mathrm{gd}(R_j|M_\mathrm{cut})\\
\Delta \xi_\mathrm{gd}(R|M_\mathrm{cut}) &= \hat{\xi}_\mathrm{gd}(R)-\xi_\mathrm{gd,model}(R|M_\mathrm{cut}),
\end{align}
where $\xi_\mathrm{gd,model}$ is our halo-model prediction with the modification of Eq.~(\ref{eq:tng_modified}) and we find $\min[\chi^2(M_\mathrm{cut})]=1.3$ with 2 degrees of freedom at $M_\mathrm{cut}=4\times10^{14}\, M_\odot$.

Our limit of $M_\mathrm{cut}$ can be translated into the constraint in the gas-to-halo mass relation at group-mass scales of $M=10^{12-13}\, M_\odot$.
Figure~\ref{fig:fgas_limit} shows our constraint of the hot-gas mass fraction defined by 
$f_\mathrm{gas,hot} < \tanh(M_{500}/[1\times 10^{12} M_\odot]) M_\mathrm{gas} / M_{500}$ as a function of halo masses.
In the figure, the orange filled region stands for our constraint by the galaxy-FRB cross correlation, 
whereas detailed mass dependence in the upper limit has little meaning.
Caution that the hot gas mass from the TNG300 depends on $M_\mathrm{500}$, leading that our limit of $f_\mathrm{gas,hot}$ exhibits a different dependence from the $\tanh$ form.
The other filled regions and pink points with error bars show
the constraints by different methods in the literture;
the cyan filled region has been derived from the stacking analysis of thermal Sunyaev--Zel'dovich (SZ) effect at galaxy groups in the Sloan Digital Sky Survey \citep{2021MNRAS.504.5131L};
the hatched region comes from multi-wavelength observations of an X-ray-selected sample \citep{2022PASJ...74..175A};
the gray filled region represents the limits inferred from a joint analysis of kinematic SZ effects of massive galaxies and cosmic-shear autocorrelations \citep{2024MNRAS.534..655B};
the pink points with error bars are the limits from the X-ray analysis in \citet{2024arXiv241116555P}.
We confirm that our limit is in good agreement with those observational results, 
although the thermal SZ limit relies on virial-temperature assumption
and the X-ray analyses may be subject to contaminations from point-source emissions. 
It is beneficial to mention that our method 
%directly probes 
\rev{is sensitive to}
electron density profiles at very small scales of $O(0.01) \mathrm{Mpc}$, whereas the current kinematic SZ analysis can not measure such small-scale signals in a direct way.

Our constraint also allows us to 
%validate
\rev{compare with} existing models of baryonic feedback effects implemented in several numerical simulations.
As representative examples, we take the predictions from the FABLE \citep{2018MNRAS.479.5385H} and its variant XFABLE \citep{2025MNRAS.542.3206H}, the FLAMINGO suite \citep{2023MNRAS.526.4978S, 2023MNRAS.526.6103K}, the BAHAMAS suite \citep{2017MNRAS.465.2936M}, and the SIMBA \citep{2019MNRAS.486.2827D}.
Different colored lines in Fig.~\ref{fig:fgas_limit} represent the predicted gas-to-halo mass relations in these models, 
%highlighting some strong feedback mechanisms may be required to deplete the hot gas at group scales.
\rev{illustrating that some feedback implementations may lead to a weaker depletion of hot gas at group scales than suggested by our constraint.}
More detailed comparisons with simulation-based predictions are highly encouraged and left for future studies.
\rev{We note that cosmological hydrodynamical simulations are typically calibrated to reproduce specific observables, such as the present-day stellar mass function, and their predictions for hot gas content can vary accordingly.}
\rev{Figure~\ref{fig:fgas_limit} therefore highlights the potential for tension with certain feedback implementations and underscores the need for additional observational constraints on gas contents in the universe to improve our understanding of cosmic structure formation.}

%\clearpage

\section{Conclusions}\label{sec:conc}

In this paper, we performed a cross-correlation analysis using the 2MASS galaxy catalog and 133 localized Fast Radio Bursts (FRBs).
By leveraging redshift information, we measured the real-space correlation function between the foreground galaxy number density and the dispersion measures of background FRBs as a function of comoving separation $R$.
The correlation function $\xi_\mathrm{gd}(R)$ was designed to be insensitive to contributions from Galactic dispersion measures and 
%nearby-source 
host-galaxy components, enabling us to probe the average free-electron density around 2MASS galaxies at an effective redshift of $\sim 0.03$.
Our main findings are summarized below:
\begin{itemize}
\item Measurements of $\xi_\mathrm{gd}(R)$ in the range $0.01 < R \, [h^{-1}\mathrm{Mpc}] < 1$ are fully consistent with a null detection at about $2\sigma$ level. We confirmed that variations in component separation for dispersion measures have minimal impact on our results. Furthermore, we found no significant systematics in our background selection, demonstrating the robustness of our estimator for $\xi_\mathrm{gd}(R)$.
\item We used the IllustrisTNG-300 simulation data to predict $\xi_\mathrm{gd}(R)$ within a standard halo-model framework. Our analysis indicates that small-scale measurements of $\xi_\mathrm{gd}(R)$ at $0.01$–$0.1\,h^{-1}\mathrm{Mpc}$ are sensitive to fluctuations in free-electron number density within group-scale halos of mass $10^{12-13}\,M_\odot$.
\item Our measured $\xi_\mathrm{gd}(R)$ at $R \sim 0.01\,h^{-1}\mathrm{Mpc}$ is incompatible with the IllustrisTNG-300-based prediction, suggesting an overestimation of hot gas masses in halos of $10^{12-13}\,M_\odot$ in the simulation. Introducing a phenomenological cutoff in the gas-to-halo mass relation improves the agreement with our measurements. We also verified that an alternative electron density model inferred from X-ray stacking analyses in \citet{2024arXiv241116555P} is consistent with our results.
\item 
\rev{Under the assumption that the halo occupation distribution of the foreground galaxies is well constrained by the previous study of \citet{2018MNRAS.473.4318A},}
comparing our measurements with the IllustrisTNG-based model allows us to place meaningful constraints on hot gas masses in halos of $10^{12-13}\,M_\odot$. We find an upper limit on the hot-gas mass fraction of $\sim 0.03$, corresponding to $\sim10\%$ of the global baryon fraction. This limit is consistent with constraints from the Sunyaev--Zel’dovich effect and X-ray emission measurements around galaxies, 
%while ruling out baryonic feedback scenarios implemented in several hydrodynamical simulation models.
\rev{while suggesting a potential tension with baryonic feedback implementations in some hydrodynamical simulation models.}
\end{itemize}
It is important to note that our measurements are based on only $O(100)$ FRBs. Ongoing and future radio surveys, combined with follow-up observations, will significantly increase the number of localized FRBs.
A larger FRB sample will reduce statistical uncertainties and enable detection of cross-correlations between large-scale structures and dispersion measures, providing a powerful tool for mapping hot gas content in the universe.
As a conservative choice, we excluded measurements at $R > 1\, h^{-1}\mathrm{Mpc}$, but detailed modeling of FRBs and cosmic dispersion measures will allow the generation of synthetic mock FRB datasets and robust evaluation of statistical uncertainties at large $R$.
Furthermore, measurements of cross-correlations across different foreground redshifts and galaxies' properties (e.g.~luminosity and colors) can precisely constrain the evolution of hot gas in large-scale structures, offering deeper insights into baryon cycles in the universe.

\section*{Acknowledgments}
%We thank *** for useful discussions. 
We thank all scientists who made their
valuable observational data of FRBs and 2MASS galaxies publicly available.
Simulation data used in this paper is kindly provided by The TNG Collaboration: \url{https://www.tng-project.org/}.

\section*{Funding}
This research is supported in part by JSPS KAKENHI grant Nos.~20H05855 (RT), 22H00130 (RT and KI), 23H04900 (KI), 23H05430 (KI), 23H01172 (KI), 24H00215 (MS and KO), 24H00221(MS), 25K17380 (KO), 25H01513 (KO), 25H00662 (KO), and 26H02045 (KI). 
MS also aknowledge research supports by JST BOOST, Japan Grant Number JPMJBY24D8.
Numerical analyses were [in part] carried out on analysis servers at the Center for Computational Astrophysics, National Astronomical Observatory of Japan.

\section*{Data availability}
The 2MASS catalogue and the IllustrisTNG-300 simulations are already publicly available.
Other data in this article will be shared on a reasonable request to the first author.

\section*{Appendix1 Cross-correlation measurements}\label{apdx:our_measurements}
 
\begin{table}
  \tbl{%\footnotemark[$*$]
  A summary of our cross correlation measurements. We measure the cross correlation $\hat{\xi}_\mathrm{gd}$ in units of 
  $\mathrm{pc}/\mathrm{cc}$
  with component separations of observed dispersion measures by adopting four different model templates (denoted as N1, Nbeta, Y1, and Ybeta). See \citet{2025arXiv251102155T} for details of the component separations.
  The error bars are estimated with 40 jackknife sampling of 2MASS galaxies.
  }{%
  \begin{tabular}{ccccc}
  $R\, [h^{-1}\mathrm{Mpc}]$ & 
  $\hat{\xi}_\mathrm{gd}$ (N1) & $\hat{\xi}_\mathrm{gd}$ (Nbeta) &
  $\hat{\xi}_\mathrm{gd}$ (Y1) & $\hat{\xi}_\mathrm{gd}$ (Ybeta) \\ \hline
  $0.0186$ & 
  $18.2\pm44.8$ & $19.2\pm44.5$ & $19.6\pm48.4$ & $20.8\pm47.5$ \\
  $0.0468$ & 
  $29.7\pm73.7$ & $39.1\pm83.1$ & $31.8\pm78.2$ & $41.7\pm88.3$ \\
  $0.117$ & 
  $48.4\pm63.9$ & $54.3\pm71.7$ & $51.4\pm69.0$ & $57.9\pm77.6$ \\
  $0.295$ & 
  $52.5\pm52.0$ & $55.9\pm56.4$ & $53.1\pm53.3$ & $56.6\pm57.9$ \\
  $0.742$ & 
  $23.6\pm26.2$ & $27.9\pm31.2$ & $23.3\pm27.4$ & $27.7\pm32.7$ \\
  \end{tabular}}\label{table:our_measurements}
  \begin{tabnote}
 %\footnotemark[$*$] A brief note of table.  
  \end{tabnote}
\end{table}
Table~\ref{table:our_measurements} provides numerical values 
of the cross correlation functions between the 2MASS galaxies and 133 localized FRBs.
The covariance matices of individual measurements are given by 
\footnotesize
\begin{equation}
\mathrm{Cov}_{ij} / \sqrt{\mathrm{Cov}_{ii}\mathrm{Cov}_{jj}} = 
\begin{pmatrix}
1 & 0.166 & 0.179 & 0.351 & 0.339\\
0.166 & 1 & 0.954 & 0.940 & 0.866\\
0.179 & 0.954 & 1 & 0.952 & 0.891\\
0.351 & 0.940 & 0.952 & 1 & 0.969\\
0.339 & 0.866 & 0.891 & 0.969 & 1\\
\end{pmatrix} \label{eq:cov_N1}
\end{equation}
\normalsize
for N1,
\footnotesize
\begin{equation}
\mathrm{Cov}_{ij} / \sqrt{\mathrm{Cov}_{ii}\mathrm{Cov}_{jj}} = 
\begin{pmatrix}
1 & 0.174 & 0.186 & 0.350 & 0.345 \\
0.174 & 1 & 0.962 & 0.943 & 0.869 \\
0.186 & 0.962 & 1 & 0.958 & 0.895 \\
0.350 & 0.943 & 0.958 & 1 & 0.972 \\
0.345 & 0.869 & 0.895 & 0.972 & 1\\
\end{pmatrix} \label{eq:cov_Nbeta}
\end{equation}
\normalsize
for Nbeta,
\footnotesize
\begin{equation}
\mathrm{Cov}_{ij} / \sqrt{\mathrm{Cov}_{ii}\mathrm{Cov}_{jj}} = 
\begin{pmatrix}
1 & 0.164 & 0.180 & 0.343 & 0.328 \\
0.164 & 1 &         0.961 & 0.950 & 0.882\\
0.180 & 0.961 & 1 & 0.956 & 0.902 \\
0.343 & 0.950 & 0.956 & 1  & 0.971 \\
0.328 & 0.882 & 0.902 & 0.971 & 1 \\
\end{pmatrix} \label{eq:cov_Y1}
\end{equation}
\normalsize
for Y1,
\footnotesize
\begin{equation}
\mathrm{Cov}_{ij} / \sqrt{\mathrm{Cov}_{ii}\mathrm{Cov}_{jj}} = 
\begin{pmatrix}
1 &        0.174 & 0.188 & 0.344 & 0.335 \\
0.174 & 1 &         0.967 & 0.951 & 0.884 \\
0.188 & 0.967 & 1 & 0.961 & 0.906\\
0.344 & 0.951 & 0.961 & 1 & 0.974\\
0.335 & 0.884 & 0.906 & 0.974 & 1\\
\end{pmatrix} \label{eq:cov_Ybeta}
\end{equation}
\normalsize
for Ybeta, respectively.

\bibliographystyle{apj}
\bibliography{refs}

\begin{thebibliography}{}
\expandafter\ifx\csname natexlab\endcsname\relax\def\natexlab#1{#1}\fi

\bibitem[{{Akino} {et~al.}(2022){Akino}, {Eckert}, {Okabe}, {Sereno}, {Umetsu}, {Oguri}, {Gastaldello}, {Chiu}, {Ettori}, {Evrard}, {Farahi}, {Maughan}, {Pierre}, {Ricci}, {Valtchanov}, {McCarthy}, {McGee}, {Miyazaki}, {Nishizawa}, \& {Tanaka}}]{2022PASJ...74..175A}
{Akino}, D., {Eckert}, D., {Okabe}, N., {et~al.} 2022, \pasj, 74, 175

\bibitem[{{Amon} \& {Efstathiou}(2022)}]{2022MNRAS.516.5355A}
{Amon}, A., \& {Efstathiou}, G. 2022, \mnras, 516, 5355

\bibitem[{{Amon} {et~al.}(2022){Amon}, {Gruen}, {Troxel}, {MacCrann}, {Dodelson}, {Choi}, {Doux}, {Secco}, {Samuroff}, {Krause}, {Cordero}, {Myles}, {DeRose}, {Wechsler}, {Gatti}, {Navarro-Alsina}, {Bernstein}, {Jain}, {Blazek}, {Alarcon}, {Fert{\'e}}, {Lemos}, {Raveri}, {Campos}, {Prat}, {S{\'a}nchez}, {Jarvis}, {Alves}, {Andrade-Oliveira}, {Baxter}, {Bechtol}, {Becker}, {Bridle}, {Camacho}, {Carnero Rosell}, {Carrasco Kind}, {Cawthon}, {Chang}, {Chen}, {Chintalapati}, {Crocce}, {Davis}, {Diehl}, {Drlica-Wagner}, {Eckert}, {Eifler}, {Elvin-Poole}, {Everett}, {Fang}, {Fosalba}, {Friedrich}, {Gaztanaga}, {Giannini}, {Gruendl}, {Harrison}, {Hartley}, {Herner}, {Huang}, {Huff}, {Huterer}, {Kuropatkin}, {Leget}, {Liddle}, {McCullough}, {Muir}, {Pandey}, {Park}, {Porredon}, {Refregier}, {Rollins}, {Roodman}, {Rosenfeld}, {Ross}, {Rykoff}, {Sanchez}, {Sevilla-Noarbe}, {Sheldon}, {Shin}, {Troja}, {Tutusaus}, {Tutusaus}, {Varga}, {Weaverdyck}, {Yanny}, {Yin}, {Zhang}, {Zuntz}, {Aguena}, {Allam}, {Annis}, {Bacon},
  {Bertin}, {Bhargava}, {Brooks}, {Buckley-Geer}, {Burke}, {Carretero}, {Costanzi}, {da Costa}, {Pereira}, {De Vicente}, {Desai}, {Dietrich}, {Doel}, {Ferrero}, {Flaugher}, {Frieman}, {Garc{\'\i}a-Bellido}, {Gaztanaga}, {Gerdes}, {Giannantonio}, {Gschwend}, {Gutierrez}, {Hinton}, {Hollowood}, {Honscheid}, {Hoyle}, {James}, {Kron}, {Kuehn}, {Lahav}, {Lima}, {Lin}, {Maia}, {Marshall}, {Martini}, {Melchior}, {Menanteau}, {Miquel}, {Mohr}, {Morgan}, {Ogando}, {Palmese}, {Paz-Chinch{\'o}n}, {Petravick}, {Pieres}, {Romer}, {Sanchez}, {Scarpine}, {Schubnell}, {Serrano}, {Smith}, {Soares-Santos}, {Tarle}, {Thomas}, {To}, {Weller}, \& {DES Collaboration}}]{2022PhRvD.105b3514A}
{Amon}, A., {Gruen}, D., {Troxel}, M.~A., {et~al.} 2022, \prd, 105, 023514

\bibitem[{{Ando} {et~al.}(2018){Ando}, {Benoit-L{\'e}vy}, \& {Komatsu}}]{2018MNRAS.473.4318A}
{Ando}, S., {Benoit-L{\'e}vy}, A., \& {Komatsu}, E. 2018, \mnras, 473, 4318

\bibitem[{{Aric{\`o}} {et~al.}(2023){Aric{\`o}}, {Angulo}, {Zennaro}, {Contreras}, {Chen}, \& {Hern{\'a}ndez-Monteagudo}}]{2023A&A...678A.109A}
{Aric{\`o}}, G., {Angulo}, R.~E., {Zennaro}, M., {et~al.} 2023, \aap, 678, A109

\bibitem[{{Asgari} {et~al.}(2021){Asgari}, {Lin}, {Joachimi}, {Giblin}, {Heymans}, {Hildebrandt}, {Kannawadi}, {St{\"o}lzner}, {Tr{\"o}ster}, {van den Busch}, {Wright}, {Bilicki}, {Blake}, {de Jong}, {Dvornik}, {Erben}, {Getman}, {Hoekstra}, {K{\"o}hlinger}, {Kuijken}, {Miller}, {Radovich}, {Schneider}, {Shan}, \& {Valentijn}}]{2021A&A...645A.104A}
{Asgari}, M., {Lin}, C.-A., {Joachimi}, B., {et~al.} 2021, \aap, 645, A104

\bibitem[{{Bigwood} {et~al.}(2025){Bigwood}, {Bourne}, {Ir{\v{s}}i{\v{c}}}, {Amon}, \& {Sijacki}}]{2025MNRAS.542.3206H}
{Bigwood}, L., {Bourne}, M.~A., {Ir{\v{s}}i{\v{c}}}, V., {Amon}, A., \& {Sijacki}, D. 2025, \mnras, 542, 3206

\bibitem[{{Bigwood} {et~al.}(2024){Bigwood}, {Amon}, {Schneider}, {Salcido}, {McCarthy}, {Preston}, {Sanchez}, {Sijacki}, {Schaan}, {Ferraro}, {Battaglia}, {Chen}, {Dodelson}, {Roodman}, {Pieres}, {Fert{\'e}}, {Alarcon}, {Drlica-Wagner}, {Choi}, {Navarro-Alsina}, {Campos}, {Ross}, {Carnero Rosell}, {Yin}, {Yanny}, {S{\'a}nchez}, {Chang}, {Davis}, {Doux}, {Gruen}, {Rykoff}, {Huff}, {Sheldon}, {Tarsitano}, {Andrade-Oliveira}, {Bernstein}, {Giannini}, {Diehl}, {Huang}, {Harrison}, {Sevilla-Noarbe}, {Tutusaus}, {Elvin-Poole}, {McCullough}, {Zuntz}, {Blazek}, {DeRose}, {Cordero}, {Prat}, {Myles}, {Eckert}, {Bechtol}, {Herner}, {Secco}, {Gatti}, {Raveri}, {Kind}, {Becker}, {Troxel}, {Jarvis}, {MacCrann}, {Friedrich}, {Alves}, {Leget}, {Chen}, {Rollins}, {Wechsler}, {Gruendl}, {Cawthon}, {Allam}, {Bridle}, {Pandey}, {Everett}, {Shin}, {Hartley}, {Fang}, {Zhang}, {Aguena}, {Annis}, {Bacon}, {Bertin}, {Bocquet}, {Brooks}, {Carretero}, {Castander}, {da Costa}, {Pereira}, {De Vicente}, {Desai}, {Doel}, {Ferrero},
  {Flaugher}, {Frieman}, {Garc{\'\i}a-Bellido}, {Gaztanaga}, {Gutierrez}, {Hinton}, {Hollowood}, {Honscheid}, {Huterer}, {James}, {Kuehn}, {Lahav}, {Lee}, {Marshall}, {Mena-Fern{\'a}ndez}, {Miquel}, {Muir}, {Paterno}, {Plazas Malag{\'o}n}, {Porredon}, {Romer}, {Samuroff}, {Sanchez}, {Sanchez Cid}, {Smith}, {Soares-Santos}, {Suchyta}, {Swanson}, {Tarle}, {To}, {Weaverdyck}, {Weller}, {Wiseman}, \& {Yamamoto}}]{2024MNRAS.534..655B}
{Bigwood}, L., {Amon}, A., {Schneider}, A., {et~al.} 2024, \mnras, 534, 655

\bibitem[{{Broxterman} {et~al.}(2025){Broxterman}, {Simon}, {Porth}, {Kuijken}, {Wright}, {Asgari}, {Bilicki}, {Heymans}, {Hildebrandt}, {Hoekstra}, {Joachimi}, {Li}, {Maturi}, {Moscardini}, {Radovich}, {Reischke}, \& {Von Wietersheim-Kramsta}}]{2025A&A...703L...3B}
{Broxterman}, J.~C., {Simon}, P., {Porth}, L., {et~al.} 2025, \aap, 703, L3

\bibitem[{{Brunner} {et~al.}(2022){Brunner}, {Liu}, {Lamer}, {Georgakakis}, {Merloni}, {Brusa}, {Bulbul}, {Dennerl}, {Friedrich}, {Liu}, {Maitra}, {Nandra}, {Ramos-Ceja}, {Sanders}, {Stewart}, {Boller}, {Buchner}, {Clerc}, {Comparat}, {Dwelly}, {Eckert}, {Finoguenov}, {Freyberg}, {Ghirardini}, {Gueguen}, {Haberl}, {Kreykenbohm}, {Krumpe}, {Osterhage}, {Pacaud}, {Predehl}, {Reiprich}, {Robrade}, {Salvato}, {Santangelo}, {Schrabback}, {Schwope}, \& {Wilms}}]{2022A&A...661A...1B}
{Brunner}, H., {Liu}, T., {Lamer}, G., {et~al.} 2022, \aap, 661, A1

\bibitem[{{Chawla} {et~al.}(2022){Chawla}, {Kaspi}, {Ransom}, {Bhardwaj}, {Boyle}, {Breitman}, {Cassanelli}, {Cubranic}, {Dong}, {Fonseca}, {Gaensler}, {Giri}, {Josephy}, {Kaczmarek}, {Leung}, {Masui}, {Mena-Parra}, {Merryfield}, {Michilli}, {M{\"u}nchmeyer}, {Ng}, {Patel}, {Pearlman}, {Petroff}, {Pleunis}, {Rahman}, {Sanghavi}, {Shin}, {Smith}, {Stairs}, \& {Tendulkar}}]{2022ApJ...927...35C}
{Chawla}, P., {Kaspi}, V.~M., {Ransom}, S.~M., {et~al.} 2022, \apj, 927, 35

\bibitem[{{Chen} {et~al.}(2023){Chen}, {Aric{\`o}}, {Huterer}, {Angulo}, {Weaverdyck}, {Friedrich}, {Secco}, {Hern{\'a}ndez-Monteagudo}, {Alarcon}, {Alves}, {Amon}, {Andrade-Oliveira}, {Baxter}, {Bechtol}, {Becker}, {Bernstein}, {Blazek}, {Brandao-Souza}, {Bridle}, {Camacho}, {Campos}, {Carnero Rosell}, {Carrasco Kind}, {Cawthon}, {Chang}, {Chen}, {Chintalapati}, {Choi}, {Cordero}, {Crocce}, {Pereira}, {Davis}, {DeRose}, {Di Valentino}, {Diehl}, {Dodelson}, {Doux}, {Drlica-Wagner}, {Eckert}, {Eifler}, {Elsner}, {Elvin-Poole}, {Everett}, {Fang}, {Fert{\'e}}, {Fosalba}, {Gatti}, {Gaztanaga}, {Giannini}, {Gruen}, {Gruendl}, {Harrison}, {Hartley}, {Herner}, {Hoffmann}, {Huang}, {Huff}, {Jain}, {Jarvis}, {Jeffrey}, {Kacprzak}, {Krause}, {Kuropatkin}, {Leget}, {Lemos}, {Liddle}, {MacCrann}, {McCullough}, {Muir}, {Myles}, {Navarro-Alsina}, {Omori}, {Pandey}, {Park}, {Porredon}, {Prat}, {Raveri}, {Refregier}, {Rollins}, {Roodman}, {Rosenfeld}, {Ross}, {Rykoff}, {Samuroff}, {S{\'a}nchez}, {Sanchez}, {Sevilla-Noarbe},
  {Sheldon}, {Shin}, {Troja}, {Troxel}, {Tutusaus}, {Varga}, {Wechsler}, {Yanny}, {Yin}, {Zhang}, {Zuntz}, {Aguena}, {Annis}, {Bacon}, {Bertin}, {Bocquet}, {Brooks}, {Burke}, {Carretero}, {Conselice}, {Costanzi}, {da Costa}, {De Vicente}, {Desai}, {Doel}, {Ferrero}, {Flaugher}, {Frieman}, {Garc{\'\i}a-Bellido}, {Gerdes}, {Giannantonio}, {Gschwend}, {Gutierrez}, {Hinton}, {Hollowood}, {Honscheid}, {James}, {Kuehn}, {Lahav}, {March}, {Marshall}, {Melchior}, {Menanteau}, {Miquel}, {Mohr}, {Morgan}, {Paz-Chinch{\'o}n}, {Pieres}, {Sanchez}, {Smith}, {Suchyta}, {Swanson}, {Tarle}, {Thomas}, {To}, \& {DES Collaboration}}]{2023MNRAS.518.5340C}
{Chen}, A., {Aric{\`o}}, G., {Huterer}, D., {et~al.} 2023, \mnras, 518, 5340

\bibitem[{{Chime/Frb Collaboration} {et~al.}(2025){Chime/Frb Collaboration}, {Amiri}, {Amouyal}, {Andersen}, {Andrew}, {Bandura}, {Bhardwaj}, {Boyle}, {Brar}, {Cassity}, {Chatterjee}, {Curtin}, {Dobbs}, {Dong}, {Dong}, {Eadie}, {Eftekhari}, {Fong}, {Fonseca}, {Gaensler}, {Halpern}, {Hessels}, {Hopkins}, {Ibik}, {Joseph}, {Kaczmarek}, {Kahinga}, {Kaspi}, {Khairy}, {Kilpatrick}, {Lanman}, {Lazda}, {Leung}, {Main}, {Mas-Ribas}, {Masui}, {McKinven}, {Mena-Parra}, {Meyers}, {Michilli}, {Milutinovic}, {Nimmo}, {Noble}, {Pandhi}, {Patil}, {Pearlman}, {Petroff}, {Pleunis}, {Prochaska}, {Rafiei-Ravandi}, {Rahman}, {Renard}, {Sammons}, {Sand}, {Scholz}, {Shah}, {Shin}, {Siegel}, {Simha}, {Smith}, {Stairs}, {Vanderlinde}, {Wang}, {Wulf}, \& {Zegmott}}]{2025ApJS..280....6C}
{Chime/Frb Collaboration}, {Amiri}, M., {Amouyal}, D., {et~al.} 2025, \apjs, 280, 6

\bibitem[{{Chisari} {et~al.}(2019){Chisari}, {Mead}, {Joudaki}, {Ferreira}, {Schneider}, {Mohr}, {Tr{\"o}ster}, {Alonso}, {McCarthy}, {Martin-Alvarez}, {Devriendt}, {Slyz}, \& {van Daalen}}]{2019OJAp....2E...4C}
{Chisari}, N.~E., {Mead}, A.~J., {Joudaki}, S., {et~al.} 2019, The Open Journal of Astrophysics, 2, 4

\bibitem[{{Cooray} \& {Sheth}(2002)}]{2002PhR...372....1C}
{Cooray}, A., \& {Sheth}, R. 2002, \physrep, 372, 1

\bibitem[{{Cordes} \& {Lazio}(2002)}]{2002astro.ph..7156C}
{Cordes}, J.~M., \& {Lazio}, T.~J.~W. 2002, arXiv e-prints, astro

\bibitem[{{Cordes} {et~al.}(2016){Cordes}, {Wharton}, {Spitler}, {Chatterjee}, \& {Wasserman}}]{2016arXiv160505890C}
{Cordes}, J.~M., {Wharton}, R.~S., {Spitler}, L.~G., {Chatterjee}, S., \& {Wasserman}, I. 2016, arXiv e-prints, arXiv:1605.05890

\bibitem[{{Dalal} {et~al.}(2023){Dalal}, {Li}, {Nicola}, {Zuntz}, {Strauss}, {Sugiyama}, {Zhang}, {Rau}, {Mandelbaum}, {Takada}, {More}, {Miyatake}, {Kannawadi}, {Shirasaki}, {Taniguchi}, {Takahashi}, {Osato}, {Hamana}, {Oguri}, {Nishizawa}, {Malag{\'o}n}, {Sunayama}, {Alonso}, {Slosar}, {Luo}, {Armstrong}, {Bosch}, {Hsieh}, {Komiyama}, {Lupton}, {Lust}, {MacArthur}, {Miyazaki}, {Murayama}, {Nishimichi}, {Okura}, {Price}, {Tait}, {Tanaka}, \& {Wang}}]{2023PhRvD.108l3519D}
{Dalal}, R., {Li}, X., {Nicola}, A., {et~al.} 2023, \prd, 108, 123519

\bibitem[{{Dav{\'e}} {et~al.}(2019){Dav{\'e}}, {Angl{\'e}s-Alc{\'a}zar}, {Narayanan}, {Li}, {Rafieferantsoa}, \& {Appleby}}]{2019MNRAS.486.2827D}
{Dav{\'e}}, R., {Angl{\'e}s-Alc{\'a}zar}, D., {Narayanan}, D., {et~al.} 2019, \mnras, 486, 2827

\bibitem[{{Dekel} \& {Silk}(1986)}]{1986ApJ...303...39D}
{Dekel}, A., \& {Silk}, J. 1986, \apj, 303, 39

\bibitem[{{Duffy} {et~al.}(2008){Duffy}, {Schaye}, {Kay}, \& {Dalla Vecchia}}]{2008MNRAS.390L..64D}
{Duffy}, A.~R., {Schaye}, J., {Kay}, S.~T., \& {Dalla Vecchia}, C. 2008, \mnras, 390, L64

\bibitem[{{Eckert} {et~al.}(2021){Eckert}, {Gaspari}, {Gastaldello}, {Le Brun}, \& {O'Sullivan}}]{2021Univ....7..142E}
{Eckert}, D., {Gaspari}, M., {Gastaldello}, F., {Le Brun}, A. M.~C., \& {O'Sullivan}, E. 2021, Universe, 7, 142

\bibitem[{{Fabian}(2012)}]{2012ARA&A..50..455F}
{Fabian}, A.~C. 2012, \araa, 50, 455

\bibitem[{{G{\'o}rski} {et~al.}(2005){G{\'o}rski}, {Hivon}, {Banday}, {Wandelt}, {Hansen}, {Reinecke}, \& {Bartelmann}}]{2005ApJ...622..759G}
{G{\'o}rski}, K.~M., {Hivon}, E., {Banday}, A.~J., {et~al.} 2005, \apj, 622, 759

\bibitem[{{Guo} \& {Lee}(2025)}]{2025MNRAS.540..289G}
{Guo}, Q., \& {Lee}, K.-G. 2025, \mnras, 540, 289

\bibitem[{{Henden} {et~al.}(2018){Henden}, {Puchwein}, {Shen}, \& {Sijacki}}]{2018MNRAS.479.5385H}
{Henden}, N.~A., {Puchwein}, E., {Shen}, S., \& {Sijacki}, D. 2018, \mnras, 479, 5385

\bibitem[{{Hirata} {et~al.}(2004){Hirata}, {Mandelbaum}, {Seljak}, {Guzik}, {Padmanabhan}, {Blake}, {Brinkmann}, {Bud{\'a}vari}, {Connolly}, {Csabai}, {Scranton}, \& {Szalay}}]{2004MNRAS.353..529H}
{Hirata}, C.~M., {Mandelbaum}, R., {Seljak}, U., {et~al.} 2004, \mnras, 353, 529

\bibitem[{{Hu} \& {Kravtsov}(2003)}]{2003ApJ...584..702H}
{Hu}, W., \& {Kravtsov}, A.~V. 2003, \apj, 584, 702

\bibitem[{{Huang} {et~al.}(2022){Huang}, {Leauthaud}, {Bradshaw}, {Hearin}, {Behroozi}, {Lange}, {Greene}, {DeRose}, {Speagle}, \& {Xhakaj}}]{2022MNRAS.515.4722H}
{Huang}, S., {Leauthaud}, A., {Bradshaw}, C., {et~al.} 2022, \mnras, 515, 4722

\bibitem[{{Huchra} {et~al.}(2012){Huchra}, {Macri}, {Masters}, {Jarrett}, {Berlind}, {Calkins}, {Crook}, {Cutri}, {Erdo{\v{g}}du}, {Falco}, {George}, {Hutcheson}, {Lahav}, {Mader}, {Mink}, {Martimbeau}, {Schneider}, {Skrutskie}, {Tokarz}, \& {Westover}}]{2012ApJS..199...26H}
{Huchra}, J.~P., {Macri}, L.~M., {Masters}, K.~L., {et~al.} 2012, \apjs, 199, 26

\bibitem[{{Hussaini} {et~al.}(2025){Hussaini}, {Connor}, {Konietzka}, {Ravi}, {Faber}, {Sharma}, \& {Sherman}}]{2025ApJ...993L..27H}
{Hussaini}, M., {Connor}, L., {Konietzka}, R.~M., {et~al.} 2025, \apjl, 993, L27

\bibitem[{{Ioka}(2003)}]{2003ApJ...598L..79I}
{Ioka}, K. 2003, \apjl, 598, L79

\bibitem[{{Jaroszy{\'n}ski}(2020)}]{2020AcA....70...87J}
{Jaroszy{\'n}ski}, M. 2020, Acta Astronomica, 70, 87

\bibitem[{{Katz} {et~al.}(1996){Katz}, {Weinberg}, \& {Hernquist}}]{1996ApJS..105...19K}
{Katz}, N., {Weinberg}, D.~H., \& {Hernquist}, L. 1996, \apjs, 105, 19

\bibitem[{{Konietzka} {et~al.}(2025){Konietzka}, {Connor}, {Semenov}, {Beane}, {Springel}, \& {Hernquist}}]{2025arXiv250707090K}
{Konietzka}, R.~M., {Connor}, L., {Semenov}, V.~A., {et~al.} 2025, arXiv e-prints, arXiv:2507.07090

\bibitem[{{Kovacs} {et~al.}(2024){Kovacs}, {Mao}, {Basu}, {Ma}, {Pakmor}, {Spitler}, \& {Walker}}]{2024A&A...690A..47K}
{Kovacs}, T.~O., {Mao}, S.~A., {Basu}, A., {et~al.} 2024, \aap, 690, A47

\bibitem[{{Kugel} {et~al.}(2023){Kugel}, {Schaye}, {Schaller}, {Helly}, {Braspenning}, {Elbers}, {Frenk}, {McCarthy}, {Kwan}, {Salcido}, {van Daalen}, {Vandenbroucke}, {Bah{\'e}}, {Borrow}, {Chaikin}, {Hu{\v{s}}ko}, {Jenkins}, {Lacey}, {Nobels}, \& {Vernon}}]{2023MNRAS.526.6103K}
{Kugel}, R., {Schaye}, J., {Schaller}, M., {et~al.} 2023, \mnras, 526, 6103

\bibitem[{{Lavaux} \& {Hudson}(2011)}]{2011MNRAS.416.2840L}
{Lavaux}, G., \& {Hudson}, M.~J. 2011, \mnras, 416, 2840

\bibitem[{{Leung} {et~al.}(2025){Leung}, {Borrow}, {Masui}, {Andrew}, {Chen}, {Schaye}, \& {Schaller}}]{2025arXiv250919514L}
{Leung}, C., {Borrow}, J., {Masui}, K.~W., {et~al.} 2025, arXiv e-prints, arXiv:2509.19514

\bibitem[{{Lewis} {et~al.}(2000){Lewis}, {Challinor}, \& {Lasenby}}]{2000ApJ...538..473L}
{Lewis}, A., {Challinor}, A., \& {Lasenby}, A. 2000, \apj, 538, 473

\bibitem[{{Li} {et~al.}(2023){Li}, {Zhang}, {Sugiyama}, {Dalal}, {Terasawa}, {Rau}, {Mandelbaum}, {Takada}, {More}, {Strauss}, {Miyatake}, {Shirasaki}, {Hamana}, {Oguri}, {Luo}, {Nishizawa}, {Takahashi}, {Nicola}, {Osato}, {Kannawadi}, {Sunayama}, {Armstrong}, {Bosch}, {Komiyama}, {Lupton}, {Lust}, {MacArthur}, {Miyazaki}, {Murayama}, {Nishimichi}, {Okura}, {Price}, {Tait}, {Tanaka}, \& {Wang}}]{2023PhRvD.108l3518L}
{Li}, X., {Zhang}, T., {Sugiyama}, S., {et~al.} 2023, \prd, 108, 123518

\bibitem[{{Lim} {et~al.}(2021){Lim}, {Barnes}, {Vogelsberger}, {Mo}, {Nelson}, {Pillepich}, {Dolag}, \& {Marinacci}}]{2021MNRAS.504.5131L}
{Lim}, S.~H., {Barnes}, D., {Vogelsberger}, M., {et~al.} 2021, \mnras, 504, 5131

\bibitem[{{Mandelbaum}(2015)}]{2015IAUS..311...86M}
{Mandelbaum}, R. 2015, in IAU Symposium, Vol. 311, Galaxy Masses as Constraints of Formation Models, ed. M.~{Cappellari} \& S.~{Courteau}, 86--95

\bibitem[{{Mandelbaum} {et~al.}(2005){Mandelbaum}, {Hirata}, {Seljak}, {Guzik}, {Padmanabhan}, {Blake}, {Blanton}, {Lupton}, \& {Brinkmann}}]{2005MNRAS.361.1287M}
{Mandelbaum}, R., {Hirata}, C.~M., {Seljak}, U., {et~al.} 2005, \mnras, 361, 1287

\bibitem[{{Marinacci} {et~al.}(2018){Marinacci}, {Vogelsberger}, {Pakmor}, {Torrey}, {Springel}, {Hernquist}, {Nelson}, {Weinberger}, {Pillepich}, {Naiman}, \& {Genel}}]{2018MNRAS.480.5113M}
{Marinacci}, F., {Vogelsberger}, M., {Pakmor}, R., {et~al.} 2018, \mnras, 480, 5113

\bibitem[{{McCarthy} {et~al.}(2017){McCarthy}, {Schaye}, {Bird}, \& {Le Brun}}]{2017MNRAS.465.2936M}
{McCarthy}, I.~G., {Schaye}, J., {Bird}, S., \& {Le Brun}, A. M.~C. 2017, \mnras, 465, 2936

\bibitem[{{Medlock} {et~al.}(2025){Medlock}, {Nagai}, {Angl{\'e}s-Alc{\'a}zar}, \& {Gebhardt}}]{2025ApJ...983...46M}
{Medlock}, I., {Nagai}, D., {Angl{\'e}s-Alc{\'a}zar}, D., \& {Gebhardt}, M. 2025, \apj, 983, 46

\bibitem[{{Medlock} {et~al.}(2024){Medlock}, {Nagai}, {Singh}, {Oppenheimer}, {Angl{\'e}s-Alc{\'a}zar}, \& {Villaescusa-Navarro}}]{2024ApJ...967...32M}
{Medlock}, I., {Nagai}, D., {Singh}, P., {et~al.} 2024, \apj, 967, 32

\bibitem[{{Merryfield} {et~al.}(2023){Merryfield}, {Tendulkar}, {Shin}, {Andersen}, {Josephy}, {Good}, {Dong}, {Masui}, {Lang}, {M{\"u}nchmeyer}, {Brar}, {Cassanelli}, {Dobbs}, {Fonseca}, {Kaspi}, {Mena-Parra}, {Pleunis}, {Rafiei-Ravandi}, {Sand}, {Scholz}, {Smith}, \& {Stairs}}]{2023AJ....165..152M}
{Merryfield}, M., {Tendulkar}, S.~P., {Shin}, K., {et~al.} 2023, \aj, 165, 152

\bibitem[{{Mo} {et~al.}(2023){Mo}, {Zhu}, {Wang}, {Tang}, \& {Feng}}]{2023MNRAS.518..539M}
{Mo}, J.-F., {Zhu}, W., {Wang}, Y., {Tang}, L., \& {Feng}, L.-L. 2023, \mnras, 518, 539

\bibitem[{{Naiman} {et~al.}(2018){Naiman}, {Pillepich}, {Springel}, {Ramirez-Ruiz}, {Torrey}, {Vogelsberger}, {Pakmor}, {Nelson}, {Marinacci}, {Hernquist}, {Weinberger}, \& {Genel}}]{2018MNRAS.477.1206N}
{Naiman}, J.~P., {Pillepich}, A., {Springel}, V., {et~al.} 2018, \mnras, 477, 1206

\bibitem[{{Nelson} {et~al.}(2018){Nelson}, {Pillepich}, {Springel}, {Weinberger}, {Hernquist}, {Pakmor}, {Genel}, {Torrey}, {Vogelsberger}, {Kauffmann}, {Marinacci}, \& {Naiman}}]{2018MNRAS.475..624N}
{Nelson}, D., {Pillepich}, A., {Springel}, V., {et~al.} 2018, \mnras, 475, 624

\bibitem[{{Nicola} {et~al.}(2022){Nicola}, {Villaescusa-Navarro}, {Spergel}, {Dunkley}, {Angl{\'e}s-Alc{\'a}zar}, {Dav{\'e}}, {Genel}, {Hernquist}, {Nagai}, {Somerville}, \& {Wandelt}}]{2022JCAP...04..046N}
{Nicola}, A., {Villaescusa-Navarro}, F., {Spergel}, D.~N., {et~al.} 2022, Journal of Cosmology and Astroparticle Physics, 2022, 046

\bibitem[{{Norberg} {et~al.}(2009){Norberg}, {Baugh}, {Gazta{\~n}aga}, \& {Croton}}]{2009MNRAS.396...19N}
{Norberg}, P., {Baugh}, C.~M., {Gazta{\~n}aga}, E., \& {Croton}, D.~J. 2009, \mnras, 396, 19

\bibitem[{{Ocker} {et~al.}(2022){Ocker}, {Cordes}, {Chatterjee}, \& {Gorsuch}}]{2022ApJ...934...71O}
{Ocker}, S.~K., {Cordes}, J.~M., {Chatterjee}, S., \& {Gorsuch}, M.~R. 2022, \apj, 934, 71

\bibitem[{{Petroff} {et~al.}(2019){Petroff}, {Hessels}, \& {Lorimer}}]{2019A&ARv..27....4P}
{Petroff}, E., {Hessels}, J.~W.~T., \& {Lorimer}, D.~R. 2019, \aapr, 27, 4

\bibitem[{{Pillepich} {et~al.}(2018){Pillepich}, {Nelson}, {Hernquist}, {Springel}, {Pakmor}, {Torrey}, {Weinberger}, {Genel}, {Naiman}, {Marinacci}, \& {Vogelsberger}}]{2018MNRAS.475..648P}
{Pillepich}, A., {Nelson}, D., {Hernquist}, L., {et~al.} 2018, \mnras, 475, 648

\bibitem[{{Planck Collaboration} {et~al.}(2020){Planck Collaboration}, {Aghanim}, {Akrami}, {Ashdown}, {Aumont}, {Baccigalupi}, {Ballardini}, {Banday}, {Barreiro}, {Bartolo}, {Basak}, {Battye}, {Benabed}, {Bernard}, {Bersanelli}, {Bielewicz}, {Bock}, {Bond}, {Borrill}, {Bouchet}, {Boulanger}, {Bucher}, {Burigana}, {Butler}, {Calabrese}, {Cardoso}, {Carron}, {Challinor}, {Chiang}, {Chluba}, {Colombo}, {Combet}, {Contreras}, {Crill}, {Cuttaia}, {de Bernardis}, {de Zotti}, {Delabrouille}, {Delouis}, {Di Valentino}, {Diego}, {Dor{\'e}}, {Douspis}, {Ducout}, {Dupac}, {Dusini}, {Efstathiou}, {Elsner}, {En{\ss}lin}, {Eriksen}, {Fantaye}, {Farhang}, {Fergusson}, {Fernandez-Cobos}, {Finelli}, {Forastieri}, {Frailis}, {Fraisse}, {Franceschi}, {Frolov}, {Galeotta}, {Galli}, {Ganga}, {G{\'e}nova-Santos}, {Gerbino}, {Ghosh}, {Gonz{\'a}lez-Nuevo}, {G{\'o}rski}, {Gratton}, {Gruppuso}, {Gudmundsson}, {Hamann}, {Handley}, {Hansen}, {Herranz}, {Hildebrandt}, {Hivon}, {Huang}, {Jaffe}, {Jones}, {Karakci}, {Keih{\"a}nen},
  {Keskitalo}, {Kiiveri}, {Kim}, {Kisner}, {Knox}, {Krachmalnicoff}, {Kunz}, {Kurki-Suonio}, {Lagache}, {Lamarre}, {Lasenby}, {Lattanzi}, {Lawrence}, {Le Jeune}, {Lemos}, {Lesgourgues}, {Levrier}, {Lewis}, {Liguori}, {Lilje}, {Lilley}, {Lindholm}, {L{\'o}pez-Caniego}, {Lubin}, {Ma}, {Mac{\'\i}as-P{\'e}rez}, {Maggio}, {Maino}, {Mandolesi}, {Mangilli}, {Marcos-Caballero}, {Maris}, {Martin}, {Martinelli}, {Mart{\'\i}nez-Gonz{\'a}lez}, {Matarrese}, {Mauri}, {McEwen}, {Meinhold}, {Melchiorri}, {Mennella}, {Migliaccio}, {Millea}, {Mitra}, {Miville-Desch{\^e}nes}, {Molinari}, {Montier}, {Morgante}, {Moss}, {Natoli}, {N{\o}rgaard-Nielsen}, {Pagano}, {Paoletti}, {Partridge}, {Patanchon}, {Peiris}, {Perrotta}, {Pettorino}, {Piacentini}, {Polastri}, {Polenta}, {Puget}, {Rachen}, {Reinecke}, {Remazeilles}, {Renzi}, {Rocha}, {Rosset}, {Roudier}, {Rubi{\~n}o-Mart{\'\i}n}, {Ruiz-Granados}, {Salvati}, {Sandri}, {Savelainen}, {Scott}, {Shellard}, {Sirignano}, {Sirri}, {Spencer}, {Sunyaev}, {Suur-Uski}, {Tauber}, {Tavagnacco},
  {Tenti}, {Toffolatti}, {Tomasi}, {Trombetti}, {Valenziano}, {Valiviita}, {Van Tent}, {Vibert}, {Vielva}, {Villa}, {Vittorio}, {Wandelt}, {Wehus}, {White}, {White}, {Zacchei}, \& {Zonca}}]{2020A&A...641A...6P}
{Planck Collaboration}, {Aghanim}, N., {Akrami}, Y., {et~al.} 2020, \aap, 641, A6

\bibitem[{{Popesso} {et~al.}(2024){Popesso}, {Biviano}, {Marini}, {Dolag}, {Vladutescu-Zopp}, {Csizi}, {Biffi}, {Lamer}, {Robothan}, {Bravo}, {Lovisari}, {Ettori}, {Angelinelli}, {Driver}, {Toptun}, {Dev}, {Mazengo}, {Merloni}, {Comparat}, {Ponti}, {Mroczkowski}, {Bulbul}, {Grandis}, \& {Bahar}}]{2024arXiv241116555P}
{Popesso}, P., {Biviano}, A., {Marini}, I., {et~al.} 2024, arXiv e-prints, arXiv:2411.16555

\bibitem[{{Reischke} \& {Hagstotz}(2025)}]{2025arXiv250717742R}
{Reischke}, R., \& {Hagstotz}, S. 2025, arXiv e-prints, arXiv:2507.17742

\bibitem[{{Reischke} {et~al.}(2025){Reischke}, {Kova{\v{c}}}, {Nicola}, {Hagstotz}, \& {Schneider}}]{2025OJAp....8E.127R}
{Reischke}, R., {Kova{\v{c}}}, M., {Nicola}, A., {Hagstotz}, S., \& {Schneider}, A. 2025, The Open Journal of Astrophysics, 8, 127

\bibitem[{{Reischke} {et~al.}(2023){Reischke}, {Neumann}, {Bertmann}, {Hagstotz}, \& {Hildebrandt}}]{2023arXiv230909766R}
{Reischke}, R., {Neumann}, D., {Bertmann}, K.~A., {Hagstotz}, S., \& {Hildebrandt}, H. 2023, arXiv e-prints, arXiv:2309.09766

\bibitem[{{Robotham} {et~al.}(2011){Robotham}, {Norberg}, {Driver}, {Baldry}, {Bamford}, {Hopkins}, {Liske}, {Loveday}, {Merson}, {Peacock}, {Brough}, {Cameron}, {Conselice}, {Croom}, {Frenk}, {Gunawardhana}, {Hill}, {Jones}, {Kelvin}, {Kuijken}, {Nichol}, {Parkinson}, {Pimbblet}, {Phillipps}, {Popescu}, {Prescott}, {Sharp}, {Sutherland}, {Taylor}, {Thomas}, {Tuffs}, {van Kampen}, \& {Wijesinghe}}]{2011MNRAS.416.2640R}
{Robotham}, A.~S.~G., {Norberg}, P., {Driver}, S.~P., {et~al.} 2011, \mnras, 416, 2640

\bibitem[{{Schaye} {et~al.}(2023){Schaye}, {Kugel}, {Schaller}, {Helly}, {Braspenning}, {Elbers}, {McCarthy}, {van Daalen}, {Vandenbroucke}, {Frenk}, {Kwan}, {Salcido}, {Bah{\'e}}, {Borrow}, {Chaikin}, {Hahn}, {Hu{\v{s}}ko}, {Jenkins}, {Lacey}, \& {Nobels}}]{2023MNRAS.526.4978S}
{Schaye}, J., {Kugel}, R., {Schaller}, M., {et~al.} 2023, \mnras, 526, 4978

\bibitem[{{Secco} {et~al.}(2022){Secco}, {Samuroff}, {Krause}, {Jain}, {Blazek}, {Raveri}, {Campos}, {Amon}, {Chen}, {Doux}, {Choi}, {Gruen}, {Bernstein}, {Chang}, {DeRose}, {Myles}, {Fert{\'e}}, {Lemos}, {Huterer}, {Prat}, {Troxel}, {MacCrann}, {Liddle}, {Kacprzak}, {Fang}, {S{\'a}nchez}, {Pandey}, {Dodelson}, {Chintalapati}, {Hoffmann}, {Alarcon}, {Alves}, {Andrade-Oliveira}, {Baxter}, {Bechtol}, {Becker}, {Brandao-Souza}, {Camacho}, {Carnero Rosell}, {Carrasco Kind}, {Cawthon}, {Cordero}, {Crocce}, {Davis}, {Di Valentino}, {Drlica-Wagner}, {Eckert}, {Eifler}, {Elidaiana}, {Elsner}, {Elvin-Poole}, {Everett}, {Fosalba}, {Friedrich}, {Gatti}, {Giannini}, {Gruendl}, {Harrison}, {Hartley}, {Herner}, {Huang}, {Huff}, {Jarvis}, {Jeffrey}, {Kuropatkin}, {Leget}, {Muir}, {Mccullough}, {Navarro Alsina}, {Omori}, {Park}, {Porredon}, {Rollins}, {Roodman}, {Rosenfeld}, {Ross}, {Rykoff}, {Sanchez}, {Sevilla-Noarbe}, {Sheldon}, {Shin}, {Troja}, {Tutusaus}, {Varga}, {Weaverdyck}, {Wechsler}, {Yanny}, {Yin}, {Zhang},
  {Zuntz}, {Abbott}, {Aguena}, {Allam}, {Annis}, {Bacon}, {Bertin}, {Bhargava}, {Bridle}, {Brooks}, {Buckley-Geer}, {Burke}, {Carretero}, {Costanzi}, {da Costa}, {De Vicente}, {Diehl}, {Dietrich}, {Doel}, {Ferrero}, {Flaugher}, {Frieman}, {Garc{\'\i}a-Bellido}, {Gaztanaga}, {Gerdes}, {Giannantonio}, {Gschwend}, {Gutierrez}, {Hinton}, {Hollowood}, {Honscheid}, {Hoyle}, {James}, {Jeltema}, {Kuehn}, {Lahav}, {Lima}, {Lin}, {Maia}, {Marshall}, {Martini}, {Melchior}, {Menanteau}, {Miquel}, {Mohr}, {Morgan}, {Ogando}, {Palmese}, {Paz-Chinch{\'o}n}, {Petravick}, {Pieres}, {Plazas Malag{\'o}n}, {Rodriguez-Monroy}, {Romer}, {Sanchez}, {Scarpine}, {Schubnell}, {Scolnic}, {Serrano}, {Smith}, {Soares-Santos}, {Suchyta}, {Swanson}, {Tarle}, {Thomas}, {To}, \& {DES Collaboration}}]{2022PhRvD.105b3515S}
{Secco}, L.~F., {Samuroff}, S., {Krause}, E., {et~al.} 2022, \prd, 105, 023515

\bibitem[{{Seebeck} {et~al.}(2021){Seebeck}, {Ravi}, {Connor}, {Law}, {Simard}, \& {Uzgil}}]{2021arXiv211207639S}
{Seebeck}, J., {Ravi}, V., {Connor}, L., {et~al.} 2021, arXiv e-prints, arXiv:2112.07639

\bibitem[{{Sharma} {et~al.}(2025{\natexlab{a}}){Sharma}, {Krause}, {Ravi}, {Reischke}, {Connor}, {Pranjal R.}, \& {Anbajagane}}]{2025arXiv250905866S}
{Sharma}, K., {Krause}, E., {Ravi}, V., {et~al.} 2025{\natexlab{a}}, arXiv e-prints, arXiv:2509.05866

\bibitem[{{Sharma} {et~al.}(2025{\natexlab{b}}){Sharma}, {Krause}, {Ravi}, {Reischke}, {R.~S.}, \& {Connor}}]{2025ApJ...989...81S}
---. 2025{\natexlab{b}}, \apj, 989, 81

\bibitem[{{Shirasaki} {et~al.}(2017{\natexlab{a}}){Shirasaki}, {Kashiyama}, \& {Yoshida}}]{2017PhRvD..95h3012S}
{Shirasaki}, M., {Kashiyama}, K., \& {Yoshida}, N. 2017{\natexlab{a}}, \prd, 95, 083012

\bibitem[{{Shirasaki} {et~al.}(2017{\natexlab{b}}){Shirasaki}, {Takada}, {Miyatake}, {Takahashi}, {Hamana}, {Nishimichi}, \& {Murata}}]{2017MNRAS.470.3476S}
{Shirasaki}, M., {Takada}, M., {Miyatake}, H., {et~al.} 2017{\natexlab{b}}, \mnras, 470, 3476

\bibitem[{{Shirasaki} {et~al.}(2022){Shirasaki}, {Takahashi}, {Osato}, \& {Ioka}}]{2022MNRAS.512.1730S}
{Shirasaki}, M., {Takahashi}, R., {Osato}, K., \& {Ioka}, K. 2022, \mnras, 512, 1730

\bibitem[{{Shivraj Patil} {et~al.}(2025){Shivraj Patil}, {Main}, {Fonseca}, {McGregor}, {Gaensler}, {Brar}, {Cook}, {Curtin}, {Eadie}, {Joseph}, {Kahinga}, {Kaspi}, {Khan}, {Kharel}, {Lanman}, {Leung}, {Masui}, {Ng}, {Nimmo}, {Pandhi}, {Pearlman}, {Pleunis}, {Sammons}, {Sand}, {Scholz}, {Shin}, {Siegel}, \& {Smith}}]{2025arXiv250906721S}
{Shivraj Patil}, S., {Main}, R.~A., {Fonseca}, E., {et~al.} 2025, arXiv e-prints, arXiv:2509.06721

\bibitem[{{Singh} {et~al.}(2017){Singh}, {Mandelbaum}, {Seljak}, {Slosar}, \& {Vazquez Gonzalez}}]{2017MNRAS.471.3827S}
{Singh}, S., {Mandelbaum}, R., {Seljak}, U., {Slosar}, A., \& {Vazquez Gonzalez}, J. 2017, \mnras, 471, 3827

\bibitem[{{Skrutskie} {et~al.}(2006){Skrutskie}, {Cutri}, {Stiening}, {Weinberg}, {Schneider}, {Carpenter}, {Beichman}, {Capps}, {Chester}, {Elias}, {Huchra}, {Liebert}, {Lonsdale}, {Monet}, {Price}, {Seitzer}, {Jarrett}, {Kirkpatrick}, {Gizis}, {Howard}, {Evans}, {Fowler}, {Fullmer}, {Hurt}, {Light}, {Kopan}, {Marsh}, {McCallon}, {Tam}, {Van Dyk}, \& {Wheelock}}]{2006AJ....131.1163S}
{Skrutskie}, M.~F., {Cutri}, R.~M., {Stiening}, R., {et~al.} 2006, \aj, 131, 1163

\bibitem[{{Springel} \& {Hernquist}(2003)}]{2003MNRAS.339..289S}
{Springel}, V., \& {Hernquist}, L. 2003, \mnras, 339, 289

\bibitem[{{Springel} {et~al.}(2018){Springel}, {Pakmor}, {Pillepich}, {Weinberger}, {Nelson}, {Hernquist}, {Vogelsberger}, {Genel}, {Torrey}, {Marinacci}, \& {Naiman}}]{2018MNRAS.475..676S}
{Springel}, V., {Pakmor}, R., {Pillepich}, A., {et~al.} 2018, \mnras, 475, 676

\bibitem[{{Takahashi} {et~al.}(2021){Takahashi}, {Ioka}, {Mori}, \& {Funahashi}}]{2021MNRAS.502.2615T}
{Takahashi}, R., {Ioka}, K., {Mori}, A., \& {Funahashi}, K. 2021, \mnras, 502, 2615

\bibitem[{{Takahashi} {et~al.}(2025){Takahashi}, {Ioka}, {Shirasaki}, \& {Osato}}]{2025arXiv251102155T}
{Takahashi}, R., {Ioka}, K., {Shirasaki}, M., \& {Osato}, K. 2025, arXiv e-prints, arXiv:2511.02155

\bibitem[{{Terasawa} {et~al.}(2025){Terasawa}, {Li}, {Takada}, {Nishimichi}, {Tanaka}, {Sugiyama}, {Kurita}, {Zhang}, {Shirasaki}, {Takahashi}, {Miyatake}, {More}, \& {Nishizawa}}]{2025PhRvD.111f3509T}
{Terasawa}, R., {Li}, X., {Takada}, M., {et~al.} 2025, \prd, 111, 063509

\bibitem[{{Theis} {et~al.}(2024){Theis}, {Hagstotz}, {Reischke}, \& {Weller}}]{2024arXiv240308611T}
{Theis}, A., {Hagstotz}, S., {Reischke}, R., \& {Weller}, J. 2024, arXiv e-prints, arXiv:2403.08611

\bibitem[{{Tinker} {et~al.}(2008){Tinker}, {Kravtsov}, {Klypin}, {Abazajian}, {Warren}, {Yepes}, {Gottl{\"o}ber}, \& {Holz}}]{2008ApJ...688..709T}
{Tinker}, J., {Kravtsov}, A.~V., {Klypin}, A., {et~al.} 2008, \apj, 688, 709

\bibitem[{{Tinker} {et~al.}(2010){Tinker}, {Robertson}, {Kravtsov}, {Klypin}, {Warren}, {Yepes}, \& {Gottl{\"o}ber}}]{2010ApJ...724..878T}
{Tinker}, J.~L., {Robertson}, B.~E., {Kravtsov}, A.~V., {et~al.} 2010, \apj, 724, 878

\bibitem[{{Umetsu}(2020)}]{2020A&ARv..28....7U}
{Umetsu}, K. 2020, \aapr, 28, 7

\bibitem[{{Valentini} \& {Dolag}(2025)}]{2025arXiv250206954V}
{Valentini}, M., \& {Dolag}, K. 2025, arXiv e-prints, arXiv:2502.06954

\bibitem[{{van den Bosch} {et~al.}(2013){van den Bosch}, {More}, {Cacciato}, {Mo}, \& {Yang}}]{2013MNRAS.430..725V}
{van den Bosch}, F.~C., {More}, S., {Cacciato}, M., {Mo}, H., \& {Yang}, X. 2013, \mnras, 430, 725

\bibitem[{{Vedantham} \& {Phinney}(2019)}]{2019MNRAS.483..971V}
{Vedantham}, H.~K., \& {Phinney}, E.~S. 2019, \mnras, 483, 971

\bibitem[{{Wang} {et~al.}(2025){Wang}, {Masui}, {Andrew}, {Fonseca}, {Gaensler}, {Joseph}, {Kaspi}, {Kharel}, {Lanman}, {Leung}, {Mas-Ribas}, {Mena-Parra}, {Nimmo}, {Pearlman}, {Pen}, {Prochaska}, {Raikman}, {Shin}, {Siegel}, {Smith}, \& {Stairs}}]{2025arXiv250608932W}
{Wang}, H., {Masui}, K., {Andrew}, S., {et~al.} 2025, arXiv e-prints, arXiv:2506.08932

\bibitem[{{Yamasaki} \& {Totani}(2020)}]{2020ApJ...888..105Y}
{Yamasaki}, S., \& {Totani}, T. 2020, \apj, 888, 105

\bibitem[{{Yao} {et~al.}(2017){Yao}, {Manchester}, \& {Wang}}]{2017ApJ...835...29Y}
{Yao}, J.~M., {Manchester}, R.~N., \& {Wang}, N. 2017, \apj, 835, 29

\bibitem[{{Zhang}(2023)}]{2023RvMP...95c5005Z}
{Zhang}, B. 2023, Reviews of Modern Physics, 95, 035005

\bibitem[{{Zhang} {et~al.}(2020){Zhang}, {Yu}, {He}, \& {Wang}}]{2020ApJ...900..170Z}
{Zhang}, G.~Q., {Yu}, H., {He}, J.~H., \& {Wang}, F.~Y. 2020, \apj, 900, 170

\bibitem[{{Zhang} {et~al.}(2025){Zhang}, {Nagamine}, {Oku}, {Lee}, {Fukushima}, {Tomaru}, {Zhang}, {Medlock}, \& {Nagai}}]{2025ApJ...993..162Z}
{Zhang}, Z.~J., {Nagamine}, K., {Oku}, Y., {et~al.} 2025, \apj, 993, 162

\bibitem[{{Zhou} {et~al.}(2014){Zhou}, {Li}, {Wang}, {Fan}, \& {Wei}}]{2014PhRvD..89j7303Z}
{Zhou}, B., {Li}, X., {Wang}, T., {Fan}, Y.-Z., \& {Wei}, D.-M. 2014, \prd, 89, 107303

\end{thebibliography}

\end{document}